\documentclass[zpreprint,zbstpl]{zeus_paper}

\usepackage[english]{babel}

\chardef\usc=95
\chardef\til=126
\catcode`\@=11 
\DeclareRobustCommand\xdotspace{\futurelet\@let@token\@xdotspace}
\def\@xdotspace{%
  \ifx\@let@token.\else
  \ifx\@let@token\bgroup.\else
  \ifx\@let@token\egroup.\else
  \ifx\@let@token\/.\else
  \ifx\@let@token\ .\else
  \ifx\@let@token~.\else
  \ifx\@let@token!.\else
  \ifx\@let@token,.\else
  \ifx\@let@token:.\else
  \ifx\@let@token;.\else
  \ifx\@let@token?.\else
  \ifx\@let@token/.\else
  \ifx\@let@token'.\else
  \ifx\@let@token).\else
  \ifx\@let@token-.\else
  \ifx\@let@token\@xobeysp.\else
  \ifx\@let@token\space.\else
  \ifx\@let@token\@sptoken.\else
   .\space
   \fi\fi\fi\fi\fi\fi\fi\fi\fi\fi\fi\fi\fi\fi\fi\fi\fi\fi}
\catcode`\@=12 

\newcommand{\stru}[2]{%
   \relax\ifmmode\hbox{\vrule height#1 depth#2 width0pt}%
   \else\vrule height#1 depth#2 width0pt\fi}

\newcommand{\Ronum}[1]{\uppercase\expandafter{\romannumeral#1}}
\newcommand{\ronum}[1]{\expandafter{\romannumeral#1}}
\DeclareRobustCommand{\LaTeXZ}{%
  \LaTeX\kern-.05em4\kern-.1em
  {\raisebox{-0.2ex}{$\scriptstyle\text{ZEUS}$}}\xspace}

\DeclareMathAlphabet{\mathbf}{OT1}{cmr}{bx}{sl}
\newcommand{\eVdist}{\kern-0.06667em}

\newcommand{\gev}{{\,\text{Ge}\eVdist\text{V\/}}}

\newcommand{\pb}{\,\text{pb}}

\newcommand{\slashfrac}[2]{%
  \raisebox{0.5ex}{\ensuremath #1}\kern-0.12em/\kern-0.08em
  \raisebox{-.8ex}{\ensuremath #2}}

\newcommand{\sqr}[3]{%
    {\vcenter{\hrule height.#3ex\hbox{\vrule width.#2ex height#1ex
     \kern#1ex\vrule width.#3ex}\hrule height.#2ex}}}

\catcode`\@=11 
\newcommand{\parenbar}{\mathpalette\p@renb@r}
\def\p@renb@r#1#2{\vbox{%
  \ifx#1\scriptscriptstyle \dimen@.7em\dimen@ii.2em\else
  \ifx#1\scriptstyle \dimen@.8em\dimen@ii.25em\else
  \dimen@1em\dimen@ii.4em\fi\fi \offinterlineskip
  \ialign{\hfill##\hfill\cr
    \vbox{\hrule width\dimen@ii}\cr
    \noalign{\vskip-.3ex}%
    \hbox to\dimen@{$\mathchar300\hfil\mathchar301$}\cr
    \noalign{\vskip-.3ex}%
    $#1#2$\cr}}}
\catcode`\@=12 

\newcommand{\IP}{{\rm I$\kern-0.01667em$P}\xspace}

\mathchardef\qsm=63
\mathchardef\pls=43
\mathchardef\mns=512
\mathchardef\plm=518
\mathchardef\eql=61
\mathchardef\smallleft=300
\mathchardef\smallright=301
\mathchardef\les=316
\mathchardef\gre=318
\mathchardef\leq=532
\mathchardef\grq=533

\catcode`\@=11 
\newcounter{pict@width}
\newcounter{pict@height}
\newlength{\pict@scale}
\newcommand{\psfigadd}[4]{%
\setcounter{pict@width}{1*\ratio{#2+\pict@scale/2}{\pict@scale}}
\setcounter{pict@height}{1*\ratio{#3+\pict@scale/2}{\pict@scale}}
\setlength{\unitlength}{\pict@scale}
\hbox to #2{\hspace{-\fill}\begin{picture}(\thepict@width,\thepict@height)
\put(0,0){\psfig{figure=#1,width=#2,height=#3,clip=}}
\SetScale{0.283466457}
\SetWidth{1.763889}
{#4}
\end{picture}}
}
\newcounter{pict@widthfst}
\newcounter{pict@widthscd}
\newcounter{pict@widthtot}
\newcommand{\psfigaddtwo}[7]{%
\setcounter{pict@widthfst}{1*\ratio{#2+\pict@scale/2}{\pict@scale}}
\setcounter{pict@widthscd}{1*\ratio{#2+#4+\pict@scale/2}{\pict@scale}}
\setcounter{pict@widthtot}{1*\ratio{#2+#4+#6+\pict@scale/2}{\pict@scale}}
\setcounter{pict@height}{1*\ratio{#3+\pict@scale/2}{\pict@scale}}
\setlength{\unitlength}{\pict@scale}
\hbox{\hspace{-\fill}\begin{picture}(\thepict@widthtot,\thepict@height)
\put(0,0){\psfig{figure=#1,width=#2,height=#3,clip=}}
\put(\thepict@widthscd,0){\psfig{figure=#5,width=#6,height=#3,clip=}}
\SetScale{0.283466457}
\SetWidth{1.763889}
{#7}
\end{picture}}
}
\newcommand{\psfigror}[4]{%
\setcounter{pict@width}{1*\ratio{#2+\pict@scale/2}{\pict@scale}}
\setcounter{pict@height}{1*\ratio{#3+\pict@scale/2}{\pict@scale}}
\setlength{\unitlength}{\pict@scale}
\hbox{\begin{picture}(\thepict@width,\thepict@height)
\put(0,\thepict@height){\psfig{figure=#1,width=#3,height=#2,clip=,angle=270}}
\SetScale{0.283466457}
\SetWidth{1.763889}
{#4}
\end{picture}}
}
\newcommand{\psfigrol}[4]{%
\setcounter{pict@width}{1*\ratio{#2+\pict@scale/2}{\pict@scale}}
\setcounter{pict@height}{1*\ratio{#3+\pict@scale/2}{\pict@scale}}
\setlength{\unitlength}{\pict@scale}
\hbox{\begin{picture}(\thepict@width,\thepict@height)
\put(0,0){\psfig{figure=#1,width=#3,height=#2,clip=,angle=90}}
\SetScale{0.283466457}
\SetWidth{1.763889}
{#4}
\end{picture}}
}
\catcode`\@=12 
\newlength\listtextwidth

\catcode`\@=11 
\newlength{\@tabfninsert}
\newlength{\@tabfnwidth}
\newcommand{\tabfootnote}[2]{%
  \setlength{\@tabfninsert}{0.8em}
  \setlength{\@tabfnwidth}{\textwidth}
  \addtolength{\@tabfnwidth}{-\@tabfninsert}
  \addtolength{\@tabfnwidth}{-0.4em}
  \noindent\makebox[\@tabfninsert][r]{\footnotesize$^{#1}$\hfil}\hfill%
  \parbox[t]{\@tabfnwidth}{\footnotesize #2\hfill}}
\catcode`\@=12 

\def\k0s{K_S^0}
\def\pb1{pb$^{-1}$}
\def\q2{Q^2}
\def\ele{e^+e^-}
\def\pp{p\bar p}
\def\g2{GeV$^2$}
\def\qq{q\bar q}
\def\as{\alpha_s}
\def\colab#1{#1 Coll.}
\def\JHEP{JHEP}

\def\figdir{./}

\begin{document}

\prepnum{{DESY--11--205}}

\title{Scaled momentum distributions\\
 for {\boldmath $\k0s$} and {\boldmath $\Lambda/\bar\Lambda$} in DIS
        at HERA}
                    
\author{ZEUS Collaboration}
\date{November 2011}

\abstract{
Scaled momentum distributions for the strange hadrons $\k0s$ and
$\Lambda/\bar\Lambda$ were measured in deep inelastic $ep$ scattering
with the ZEUS detector at HERA using an integrated luminosity of 330~\pb1. 
The evolution of these distributions with the photon virtuality,
$\q2$, was studied in the kinematic region $10<\q2<40000$~\g2 and
$0.001<x<0.75$, where $x$ is the Bjorken scaling variable. Clear
scaling violations are observed. Predictions based on different
approaches to fragmentation were compared to the measurements.
Leading-logarithm parton-shower Monte Carlo calculations interfaced to
the Lund string fragmentation model describe the data reasonably well
in the whole range measured. Next-to-leading-order QCD calculations 
based on fragmentation functions, FFs, extracted from $\ele$ data alone,
fail to describe the measurements. The calculations based on FFs
extracted from a global analysis including $\ele$, $ep$ and $pp$ data
give an improved description. The measurements presented in this paper
have the potential to further constrain the FFs of quarks, anti-quarks
and gluons yielding $\k0s$ and $\Lambda/\bar\Lambda$ strange hadrons.
}

\makezeustitle

\pagenumbering{Roman}
\begin{center}
{                      \Large  The ZEUS Collaboration              }
\end{center}

{\small
        {\raggedright
H.~Abramowicz$^{45, ah}$, 
I.~Abt$^{35}$, 
L.~Adamczyk$^{13}$, 
M.~Adamus$^{54}$, 
R.~Aggarwal$^{7, c}$, 
S.~Antonelli$^{4}$, 
P.~Antonioli$^{3}$, 
A.~Antonov$^{33}$, 
M.~Arneodo$^{50}$, 
V.~Aushev$^{26, 27, z}$
% ...
        }

\newpage

\makebox[3em]{$^{1}$}
\begin{minipage}[t]{14cm}
{\it Argonne National Laboratory, Argonne, Illinois 60439-4815, USA}~$^{A}$

\end{minipage}\\
\makebox[3em]{$^{2}$}
\begin{minipage}[t]{14cm}
{\it Andrews University, Berrien Springs, Michigan 49104-0380, USA}

\end{minipage}\\
\makebox[3em]{$^{3}$}
\begin{minipage}[t]{14cm}
{\it INFN Bologna, Bologna, Italy}~$^{B}$

\end{minipage}\\
\makebox[3em]{$^{4}$}
\begin{minipage}[t]{14cm}
{\it University and INFN Bologna, Bologna, Italy}~$^{B}$

\end{minipage}\\
\makebox[3em]{$^{5}$}
\begin{minipage}[t]{14cm}
{\it Physikalisches Institut der Universit\"at Bonn,
Bonn, Germany}~$^{C}$

\end{minipage}\\
\makebox[3em]{$^{6}$}
\begin{minipage}[t]{14cm}
{\it H.H.~Wills Physics Laboratory, University of Bristol,
Bristol, United Kingdom}~$^{D}$

\end{minipage}\\
\makebox[3em]{$^{7}$}
\begin{minipage}[t]{14cm}
{\it Panjab University, Department of Physics, Chandigarh, India}

\end{minipage}\\
\makebox[3em]{$^{8}$}
\begin{minipage}[t]{14cm}
{\it Calabria University,
Physics Department and INFN, Cosenza, Italy}~$^{B}$

\end{minipage}\\
\makebox[3em]{$^{9}$}
\begin{minipage}[t]{14cm}
{\it Institute for Universe and Elementary Particles, Chonnam National University,\\
Kwangju, South Korea}

\end{minipage}\\
\makebox[3em]{$^{10}$}
\begin{minipage}[t]{14cm}
{\it Jabatan Fizik, Universiti Malaya, 50603 Kuala Lumpur, Malaysia}~$^{E}$

\end{minipage}\\
\makebox[3em]{$^{11}$}
\begin{minipage}[t]{14cm}
{\it Nevis Laboratories, Columbia University, Irvington on Hudson,
New York 10027, USA}~$^{F}$

\end{minipage}\\
\makebox[3em]{$^{12}$}
\begin{minipage}[t]{14cm}
{\it The Henryk Niewodniczanski Institute of Nuclear Physics, Polish Academy of \\
Sciences, Krakow, Poland}~$^{G}$

\end{minipage}\\
\makebox[3em]{$^{13}$}
\begin{minipage}[t]{14cm}
{\it AGH-University of Science and Technology, Faculty of Physics and Applied Computer
Science, Krakow, Poland}~$^{H}$

\end{minipage}\\
\makebox[3em]{$^{14}$}
\begin{minipage}[t]{14cm}
{\it Department of Physics, Jagellonian University, Cracow, Poland}

\end{minipage}\\
\makebox[3em]{$^{15}$}
\begin{minipage}[t]{14cm}
{\it Deutsches Elektronen-Synchrotron DESY, Hamburg, Germany}

\end{minipage}\\
\makebox[3em]{$^{16}$}
\begin{minipage}[t]{14cm}
{\it Deutsches Elektronen-Synchrotron DESY, Zeuthen, Germany}

\end{minipage}\\
\makebox[3em]{$^{17}$}
\begin{minipage}[t]{14cm}
{\it INFN Florence, Florence, Italy}~$^{B}$

\end{minipage}\\
\makebox[3em]{$^{18}$}
\begin{minipage}[t]{14cm}
{\it University and INFN Florence, Florence, Italy}~$^{B}$

\end{minipage}\\
\makebox[3em]{$^{19}$}
\begin{minipage}[t]{14cm}
{\it Fakult\"at f\"ur Physik der Universit\"at Freiburg i.Br.,
Freiburg i.Br., Germany}

\end{minipage}\\
\makebox[3em]{$^{20}$}
\begin{minipage}[t]{14cm}
{\it School of Physics and Astronomy, University of Glasgow,
Glasgow, United Kingdom}~$^{D}$

\end{minipage}\\
\makebox[3em]{$^{21}$}
\begin{minipage}[t]{14cm}
{\it Department of Engineering in Management and Finance, Univ. of
the Aegean, Chios, Greece}

\end{minipage}\\
\makebox[3em]{$^{22}$}
\begin{minipage}[t]{14cm}
{\it Hamburg University, Institute of Experimental Physics, Hamburg,
Germany}~$^{I}$

\end{minipage}\\
\makebox[3em]{$^{23}$}
\begin{minipage}[t]{14cm}
{\it Imperial College London, High Energy Nuclear Physics Group,
London, United Kingdom}~$^{D}$

\end{minipage}\\
\makebox[3em]{$^{24}$}
\begin{minipage}[t]{14cm}
{\it Institute of Particle and Nuclear Studies, KEK,
Tsukuba, Japan}~$^{J}$

\end{minipage}\\
\makebox[3em]{$^{25}$}
\begin{minipage}[t]{14cm}
{\it Institute of Physics and Technology of Ministry of Education and
Science of Kazakhstan, Almaty, Kazakhstan}

\end{minipage}\\
\makebox[3em]{$^{26}$}
\begin{minipage}[t]{14cm}
{\it Institute for Nuclear Research, National Academy of Sciences, Kyiv, Ukraine}

\end{minipage}\\
\makebox[3em]{$^{27}$}
\begin{minipage}[t]{14cm}
{\it Department of Nuclear Physics, National Taras Shevchenko University of Kyiv, Kyiv, Ukraine}

\end{minipage}\\
\makebox[3em]{$^{28}$}
\begin{minipage}[t]{14cm}
{\it Kyungpook National University, Center for High Energy Physics, Daegu,
South Korea}~$^{K}$

\end{minipage}\\
\makebox[3em]{$^{29}$}
\begin{minipage}[t]{14cm}
{\it Institut de Physique Nucl\'{e}aire, Universit\'{e} Catholique de Louvain, Louvain-la-Neuve,\\
Belgium}~$^{L}$

\end{minipage}\\
\makebox[3em]{$^{30}$}
\begin{minipage}[t]{14cm}
{\it Departamento de F\'{\i}sica Te\'orica, Universidad Aut\'onoma
de Madrid, Madrid, Spain}~$^{M}$

\end{minipage}\\
\makebox[3em]{$^{31}$}
\begin{minipage}[t]{14cm}
{\it Department of Physics, McGill University,
Montr\'eal, Qu\'ebec, Canada H3A 2T8}~$^{N}$

\end{minipage}\\
\makebox[3em]{$^{32}$}
\begin{minipage}[t]{14cm}
{\it Meiji Gakuin University, Faculty of General Education,
Yokohama, Japan}~$^{J}$

\end{minipage}\\
\makebox[3em]{$^{33}$}
\begin{minipage}[t]{14cm}
{\it Moscow Engineering Physics Institute, Moscow, Russia}~$^{O}$

\end{minipage}\\
\makebox[3em]{$^{34}$}
\begin{minipage}[t]{14cm}
{\it Moscow State University, Institute of Nuclear Physics,
Moscow, Russia}~$^{P}$

\end{minipage}\\
\makebox[3em]{$^{35}$}
\begin{minipage}[t]{14cm}
{\it Max-Planck-Institut f\"ur Physik, M\"unchen, Germany}

\end{minipage}\\
\makebox[3em]{$^{36}$}
\begin{minipage}[t]{14cm}
{\it NIKHEF and University of Amsterdam, Amsterdam, Netherlands}~$^{Q}$

\end{minipage}\\
\makebox[3em]{$^{37}$}
\begin{minipage}[t]{14cm}
{\it Physics Department, Ohio State University,
Columbus, Ohio 43210, USA}~$^{A}$

\end{minipage}\\
\makebox[3em]{$^{38}$}
\begin{minipage}[t]{14cm}
{\it Department of Physics, University of Oxford,
Oxford, United Kingdom}~$^{D}$

\end{minipage}\\
\makebox[3em]{$^{39}$}
\begin{minipage}[t]{14cm}
{\it INFN Padova, Padova, Italy}~$^{B}$

\end{minipage}\\
\makebox[3em]{$^{40}$}
\begin{minipage}[t]{14cm}
{\it Dipartimento di Fisica dell' Universit\`a and INFN,
Padova, Italy}~$^{B}$

\end{minipage}\\
\makebox[3em]{$^{41}$}
\begin{minipage}[t]{14cm}
{\it Department of Physics, Pennsylvania State University, University Park,\\
Pennsylvania 16802, USA}~$^{F}$

\end{minipage}\\
\makebox[3em]{$^{42}$}
\begin{minipage}[t]{14cm}
{\it Polytechnic University, Sagamihara, Japan}~$^{J}$

\end{minipage}\\
\makebox[3em]{$^{43}$}
\begin{minipage}[t]{14cm}
{\it Dipartimento di Fisica, Universit\`a 'La Sapienza' and INFN,
Rome, Italy}~$^{B}$

\end{minipage}\\
\makebox[3em]{$^{44}$}
\begin{minipage}[t]{14cm}
{\it Rutherford Appleton Laboratory, Chilton, Didcot, Oxon,
United Kingdom}~$^{D}$

\end{minipage}\\
\makebox[3em]{$^{45}$}
\begin{minipage}[t]{14cm}
{\it Raymond and Beverly Sackler Faculty of Exact Sciences, School of Physics, \\
Tel Aviv University, Tel Aviv, Israel}~$^{R}$

\end{minipage}\\
\makebox[3em]{$^{46}$}
\begin{minipage}[t]{14cm}
{\it Department of Physics, Tokyo Institute of Technology,
Tokyo, Japan}~$^{J}$

\end{minipage}\\
\makebox[3em]{$^{47}$}
\begin{minipage}[t]{14cm}
{\it Department of Physics, University of Tokyo,
Tokyo, Japan}~$^{J}$

\end{minipage}\\
\makebox[3em]{$^{48}$}
\begin{minipage}[t]{14cm}
{\it Tokyo Metropolitan University, Department of Physics,
Tokyo, Japan}~$^{J}$

\end{minipage}\\
\makebox[3em]{$^{49}$}
\begin{minipage}[t]{14cm}
{\it Universit\`a di Torino and INFN, Torino, Italy}~$^{B}$

\end{minipage}\\
\makebox[3em]{$^{50}$}
\begin{minipage}[t]{14cm}
{\it Universit\`a del Piemonte Orientale, Novara, and INFN, Torino,
Italy}~$^{B}$

\end{minipage}\\
\makebox[3em]{$^{51}$}
\begin{minipage}[t]{14cm}
{\it Department of Physics, University of Toronto, Toronto, Ontario,
Canada M5S 1A7}~$^{N}$

\end{minipage}\\
\makebox[3em]{$^{52}$}
\begin{minipage}[t]{14cm}
{\it Physics and Astronomy Department, University College London,
London, United Kingdom}~$^{D}$

\end{minipage}\\
\makebox[3em]{$^{53}$}
\begin{minipage}[t]{14cm}
{\it Faculty of Physics, University of Warsaw, Warsaw, Poland}

\end{minipage}\\
\makebox[3em]{$^{54}$}
\begin{minipage}[t]{14cm}
{\it National Centre for Nuclear Research, Warsaw, Poland}

\end{minipage}\\
\makebox[3em]{$^{55}$}
\begin{minipage}[t]{14cm}
{\it Department of Particle Physics and Astrophysics, Weizmann
Institute, Rehovot, Israel}

\end{minipage}\\
\makebox[3em]{$^{56}$}
\begin{minipage}[t]{14cm}
{\it Department of Physics, University of Wisconsin, Madison,
Wisconsin 53706, USA}~$^{A}$

\end{minipage}\\
\makebox[3em]{$^{57}$}
\begin{minipage}[t]{14cm}
{\it Department of Physics, York University, Ontario, Canada M3J
1P3}~$^{N}$

\end{minipage}\\
\vspace{30em} \pagebreak[4]

\makebox[3ex]{$^{ A}$}
\begin{minipage}[t]{14cm}
 supported by the US Department of Energy\
\end{minipage}\\
\makebox[3ex]{$^{ B}$}
\begin{minipage}[t]{14cm}
 supported by the Italian National Institute for Nuclear Physics (INFN) \
\end{minipage}\\
\makebox[3ex]{$^{ C}$}
\begin{minipage}[t]{14cm}
 supported by the German Federal Ministry for Education and Research (BMBF), under
 contract No. 05 H09PDF\
\end{minipage}\\
\makebox[3ex]{$^{ D}$}
\begin{minipage}[t]{14cm}
 supported by the Science and Technology Facilities Council, UK\
\end{minipage}\\
\makebox[3ex]{$^{ E}$}
\begin{minipage}[t]{14cm}
 supported by an FRGS grant from the Malaysian government\
\end{minipage}\\
\makebox[3ex]{$^{ F}$}
\begin{minipage}[t]{14cm}
 supported by the US National Science Foundation. Any opinion,
 findings and conclusions or recommendations expressed in this material
 are those of the authors and do not necessarily reflect the views of the
 National Science Foundation.\
\end{minipage}\\
\makebox[3ex]{$^{ G}$}
\begin{minipage}[t]{14cm}
 supported by the Polish Ministry of Science and Higher Education as a scientific project No.
 DPN/N188/DESY/2009\
\end{minipage}\\
\makebox[3ex]{$^{ H}$}
\begin{minipage}[t]{14cm}
 supported by the Polish Ministry of Science and Higher Education and its grants
 for Scientific Research\
\end{minipage}\\
\makebox[3ex]{$^{ I}$}
\begin{minipage}[t]{14cm}
 supported by the German Federal Ministry for Education and Research (BMBF), under
 contract No. 05h09GUF, and the SFB 676 of the Deutsche Forschungsgemeinschaft (DFG) \
\end{minipage}\\
\makebox[3ex]{$^{ J}$}
\begin{minipage}[t]{14cm}
 supported by the Japanese Ministry of Education, Culture, Sports, Science and Technology
 (MEXT) and its grants for Scientific Research\
\end{minipage}\\
\makebox[3ex]{$^{ K}$}
\begin{minipage}[t]{14cm}
 supported by the Korean Ministry of Education and Korea Science and Engineering
 Foundation\
\end{minipage}\\
\makebox[3ex]{$^{ L}$}
\begin{minipage}[t]{14cm}
 supported by FNRS and its associated funds (IISN and FRIA) and by an Inter-University
 Attraction Poles Programme subsidised by the Belgian Federal Science Policy Office\
\end{minipage}\\
\makebox[3ex]{$^{ M}$}
\begin{minipage}[t]{14cm}
 supported by the Spanish Ministry of Education and Science through funds provided by
 CICYT\
\end{minipage}\\
\makebox[3ex]{$^{ N}$}
\begin{minipage}[t]{14cm}
 supported by the Natural Sciences and Engineering Research Council of Canada (NSERC) \
\end{minipage}\\
\makebox[3ex]{$^{ O}$}
\begin{minipage}[t]{14cm}
 partially supported by the German Federal Ministry for Education and Research (BMBF)\
\end{minipage}\\
\makebox[3ex]{$^{ P}$}
\begin{minipage}[t]{14cm}
 supported by RF Presidential grant N 4142.2010.2 for Leading Scientific Schools, by the
 Russian Ministry of Education and Science through its grant for Scientific Research on
 High Energy Physics and under contract No.02.740.11.0244 \
\end{minipage}\\
\makebox[3ex]{$^{ Q}$}
\begin{minipage}[t]{14cm}
 supported by the Netherlands Foundation for Research on Matter (FOM)\
\end{minipage}\\
\makebox[3ex]{$^{ R}$}
\begin{minipage}[t]{14cm}
 supported by the Israel Science Foundation\
\end{minipage}\\
\vspace{30em} \pagebreak[4]

\makebox[3ex]{$^{ a}$}
\begin{minipage}[t]{14cm}
now at University of Salerno, Italy\
\end{minipage}\\
\makebox[3ex]{$^{ b}$}
\begin{minipage}[t]{14cm}
now at Queen Mary University of London, United Kingdom\
\end{minipage}\\
\makebox[3ex]{$^{ c}$}
\begin{minipage}[t]{14cm}
also funded by Max Planck Institute for Physics, Munich, Germany\
\end{minipage}\\
\makebox[3ex]{$^{ d}$}
\begin{minipage}[t]{14cm}
also Senior Alexander von Humboldt Research Fellow at Hamburg University,
 Institute of Experimental Physics, Hamburg, Germany\
\end{minipage}\\
\makebox[3ex]{$^{ e}$}
\begin{minipage}[t]{14cm}
also at Cracow University of Technology, Faculty of Physics,
 Mathemathics and Applied Computer Science, Poland\
\end{minipage}\\
\makebox[3ex]{$^{ f}$}
\begin{minipage}[t]{14cm}
supported by the research grant No. 1 P03B 04529 (2005-2008)\
\end{minipage}\\
\makebox[3ex]{$^{ g}$}
\begin{minipage}[t]{14cm}
now at Rockefeller University, New York, NY
 10065, USA\
\end{minipage}\\
\makebox[3ex]{$^{ h}$}
\begin{minipage}[t]{14cm}
now at DESY group FS-CFEL-1\
\end{minipage}\\
\makebox[3ex]{$^{ i}$}
\begin{minipage}[t]{14cm}
now at Institute of High Energy Physics, Beijing, China\
\end{minipage}\\
\makebox[3ex]{$^{ j}$}
\begin{minipage}[t]{14cm}
now at DESY group FEB, Hamburg, Germany\
\end{minipage}\\
\makebox[3ex]{$^{ k}$}
\begin{minipage}[t]{14cm}
also at Moscow State University, Russia\
\end{minipage}\\
\makebox[3ex]{$^{ l}$}
\begin{minipage}[t]{14cm}
now at University of Liverpool, United Kingdom\
\end{minipage}\\
\makebox[3ex]{$^{ m}$}
\begin{minipage}[t]{14cm}
now at CERN, Geneva, Switzerland\
\end{minipage}\\
\makebox[3ex]{$^{ n}$}
\begin{minipage}[t]{14cm}
also affiliated with Universtiy College London, UK\
\end{minipage}\\
\makebox[3ex]{$^{ o}$}
\begin{minipage}[t]{14cm}
now at Goldman Sachs, London, UK\
\end{minipage}\\
\makebox[3ex]{$^{ p}$}
\begin{minipage}[t]{14cm}
also at Institute of Theoretical and Experimental Physics, Moscow, Russia\
\end{minipage}\\
\makebox[3ex]{$^{ q}$}
\begin{minipage}[t]{14cm}
also at FPACS, AGH-UST, Cracow, Poland\
\end{minipage}\\
\makebox[3ex]{$^{ r}$}
\begin{minipage}[t]{14cm}
partially supported by Warsaw University, Poland\
\end{minipage}\\
\makebox[3ex]{$^{ s}$}
\begin{minipage}[t]{14cm}
now at Istituto Nucleare di Fisica Nazionale (INFN), Pisa, Italy\
\end{minipage}\\
\makebox[3ex]{$^{ t}$}
\begin{minipage}[t]{14cm}
now at Haase Energie Technik AG, Neum\"unster, Germany\
\end{minipage}\\
\makebox[3ex]{$^{ u}$}
\begin{minipage}[t]{14cm}
now at Department of Physics, University of Bonn, Germany\
\end{minipage}\\
\makebox[3ex]{$^{ v}$}
\begin{minipage}[t]{14cm}
now at Biodiversit\"at und Klimaforschungszentrum (BiK-F), Frankfurt, Germany\
\end{minipage}\\
\makebox[3ex]{$^{ w}$}
\begin{minipage}[t]{14cm}
also affiliated with DESY, Germany\
\end{minipage}\\
\makebox[3ex]{$^{ x}$}
\begin{minipage}[t]{14cm}
also at University of Tokyo, Japan\
\end{minipage}\\
\makebox[3ex]{$^{ y}$}
\begin{minipage}[t]{14cm}
now at Kobe University, Japan\
\end{minipage}\\
\makebox[3ex]{$^{ z}$}
\begin{minipage}[t]{14cm}
supported by DESY, Germany\
\end{minipage}\\
\makebox[3ex]{$^{\dagger}$}
\begin{minipage}[t]{14cm}
 deceased \
\end{minipage}\\
\makebox[3ex]{$^{aa}$}
\begin{minipage}[t]{14cm}
member of National Technical University of Ukraine, Kyiv Polytechnic Institute,
 Kyiv, Ukraine\
\end{minipage}\\
\makebox[3ex]{$^{ab}$}
\begin{minipage}[t]{14cm}
member of National University of Kyiv - Mohyla Academy, Kyiv, Ukraine\
\end{minipage}\\
\makebox[3ex]{$^{ac}$}
\begin{minipage}[t]{14cm}
partly supported by the Russian Foundation for Basic Research, grant 11-02-91345-DFG\_a\
\end{minipage}\\
\makebox[3ex]{$^{ad}$}
\begin{minipage}[t]{14cm}
Alexander von Humboldt Professor; also at DESY and University of
 Oxford\
\end{minipage}\\
\makebox[3ex]{$^{ae}$}
\begin{minipage}[t]{14cm}
STFC Advanced Fellow\
\end{minipage}\\
\makebox[3ex]{$^{af}$}
\begin{minipage}[t]{14cm}
nee Korcsak-Gorzo\
\end{minipage}\\
\makebox[3ex]{$^{ag}$}
\begin{minipage}[t]{14cm}
This material was based on work supported by the
 National Science Foundation, while working at the Foundation.\
\end{minipage}\\
\makebox[3ex]{$^{ah}$}
\begin{minipage}[t]{14cm}
also at Max Planck Institute for Physics, Munich, Germany, External Scientific Member\
\end{minipage}\\
\makebox[3ex]{$^{ai}$}
\begin{minipage}[t]{14cm}
now at Tokyo Metropolitan University, Japan\
\end{minipage}\\
\makebox[3ex]{$^{aj}$}
\begin{minipage}[t]{14cm}
now at Nihon Institute of Medical Science, Japan\
\end{minipage}\\
\makebox[3ex]{$^{ak}$}
\begin{minipage}[t]{14cm}
now at Osaka University, Osaka, Japan\
\end{minipage}\\
\makebox[3ex]{$^{al}$}
\begin{minipage}[t]{14cm}
also at \L\'{o}d\'{z} University, Poland\
\end{minipage}\\
\makebox[3ex]{$^{am}$}
\begin{minipage}[t]{14cm}
member of \L\'{o}d\'{z} University, Poland\
\end{minipage}\\
\makebox[3ex]{$^{an}$}
\begin{minipage}[t]{14cm}
now at Department of Physics, Stockholm University, Stockholm, Sweden\
\end{minipage}\\
\makebox[3ex]{$^{ao}$}
\begin{minipage}[t]{14cm}
also at Cardinal Stefan Wyszy\'nski University, Warsaw, Poland\
\end{minipage}\\

\newpage
\pagenumbering{arabic} 
\pagestyle{plain}

\section{Introduction}
\label{intro}
The jet fragmentation and hadronisation processes through which
coloured partons become bound in colour-neutral hadrons cannot be
described within the framework of perturbative QCD (pQCD). Several
approaches have been developed which attempt to build a bridge between
the fixed-order partonic cross sections and the observed hadrons. Two of the
most successful and widely used approaches are the Lund string
model~\cite{prep:97:31} and the fragmentation functions
(FFs)~\cite{np:b160:301,zfp:c11:293,np:b421:473,np:b193:381,np:b194:445}.
The Lund string model, relying on a large number of parameters,
is interfaced to leading-logarithm parton-shower Monte Carlo
models. The FFs are parameterisations of the hadronisation process
within the standard framework of leading-twist collinear QCD
factorisation, in a similar way to that of the parton distribution
functions (PDFs), and are convoluted with the predicted partonic cross
sections.

Extensive studies of the fragmentation properties of the hadronic
final state have been performed in
$\ele$~\cite{prep:294:1,epj:c35:2004,pr:d41:2675,epj:c5:585,pl:b459:397,epj:c18:203,pl:b643:147,prep:399:71,pr:d37:1,zfp:c72:191,epj:c16:407,zfp:c47:187,pl:b311:408},
$pp$~\cite{prl:98:252001,prl:91:241803,prl:97:152302,pr:c75:064901},
$\pp$~\cite{pr:d72:052001}
and deep inelastic $ep$ scattering\footnote{Here and in the following,
  the term ``electron'' and the symbol ``$e$'' denote generically both
  the electron ($e^-$) and the positron ($e^+$), unless otherwise
  stated.}
(DIS)~\cite{zfp:c67:93,pl:b414:428,epj:c11:251,jhep:1006:009,pl:b654:148,np:b445:3,np:b504:125,pl:b681:125,epj:c61:185}
data and have provided information about the fragmentation and
hadronisation processes. The measurements provided tests of pQCD and
showed that scaling violations are observed. In addition, the
comparison of the measurements in different reactions indicated an
approximately universal behaviour of quark fragmentation.

In a previous publication~\cite{jhep:1006:009}, the ZEUS Collaboration
presented high-precision measurements of inclusive charged-hadron
production. Next-to-leading-order (NLO) QCD calculations, based on
different FFs obtained from
fits~\cite{pr:d62:054001,prl:85:5288,pr:d75:034018} to $\ele$ data, from
fits~\cite{np:b803:42} to $\ele$, $pp$ and $\pp$ data and from
fits~\cite{pr:d75:114010,pr:d76:074033} to $\ele$, $pp$ and $ep$ data,
were compared to the measurements. The predictions based on the
different FFs are similar and fail to provide a good description of
the measurements over the full range of applicability of the
calculations.
The parameterisations~\cite{pr:d75:114010,np:b734:50,epj:c61:603} of the
FFs for strange hadrons, such as $\k0s$ and $\Lambda$, are so far largely 
unconstrained. The $ep$ data presented in this paper have the
potential to constrain these FFs over a wide kinematic range.

In this paper, the scaled momentum distributions for $\k0s$ and
$\Lambda$ hadrons\footnote{Here and in the following, the notation
  $\Lambda$ includes both the particle and its antiparticle unless
  otherwise stated.} are presented for the first time in DIS. The
scaled momentum is defined as $x_p=2P^{\rm Breit}/\sqrt{\q2}$, where 
$P^{\rm Breit}$ is the particle momentum in the Breit frame and
$\q2$ is the photon virtuality. The Breit
frame~\cite{bookfeynam:1972,zfp:c2:237} is the frame in which the
exchanged virtual boson is purely space-like, with 3-momentum ${\bf
  q}=(0,0,-Q)$, providing a maximal separation between the products
of the beam fragmentation and the hard interaction. The measurements
were performed in the current region of the Breit frame, which is
equivalent to one hemisphere in $\ele$ annihilations, as functions of
$\q2$ and $x_p$. Next-to-leading-order predictions, based on different
FFs, and leading-logarithm parton-shower Monte Carlo calculations,
interfaced with the Lund string fragmentation model, were compared to
the measurements.

\section{Theoretical framework}
\label{theory}
In lowest-order QCD, three processes contribute to the DIS cross
section, namely the Born ($V^{*} q\rightarrow q$, with $V^*=\gamma^*, Z^*$),
the boson-gluon-fusion ($V^{*} g\rightarrow \qq$) and 
QCD-Compton-scattering ($V^{*} q \rightarrow qg$) processes.
The cross section for the production of an observed hadron, $H$, in the
final state in DIS can be expressed in QCD, using the factorisation
theorem, as

$$\sigma(ep\rightarrow e+H+X)=\sum_{j,j^{\prime}=q,\bar q,g} 
f_{j/p}(x,Q)\otimes \hat\sigma_{jj^{\prime}}(x,Q,z)\otimes F_{H/j^{\prime}}(z,Q),$$
where the sum runs over all possible initial (final)-state partons $j$
($j^{\prime}$), $f_{j/p}$ are the proton PDFs, which give the
probability of finding a parton $j$ with momentum fraction $x$ in the
proton, $\hat\sigma_{jj^{\prime}}$ is the partonic cross section,
which includes the matrix elements for the three processes mentioned
above, and $F_{H/j^{\prime}}$ are the FFs, which give the probability
that a hadron $H$ with momentum fraction $z$ originates from parton
$j^{\prime}$. The scaled momentum variable $x_p$ is an estimator of
$z$. As for the PDFs, the FFs include contributions from quark,
anti-quark and gluon fragmentation. Absolute predictions for the FFs
cannot be calculated; however, the dependence of the FFs on the scale
$Q$ is calculable in pQCD and governed by renormalisation group equations,
similar as for the PDFs.

The range of applicability of the FFs is limited to medium to large
values of $z$, since the assumption of massless hadrons leads to a
strong singular behaviour for $z\rightarrow 0$. At small $z$, finite
mass corrections are important. However, the inclusion of small-$z$ mass
corrections is not compatible with the factorisation theorem and thus
the FFs with mass corrections cannot be used with fixed-order
calculations. A possible solution is to introduce {\it a posteriori}
mass-correction factors to take this effect into
account~\cite{np:b803:42}.

A large improvement in the precision of the ingredients of the
calculations has been achieved in the last few years. Matrix elements
up to NLO accuracy are available for many processes; for DIS, this
corresponds to ${\cal O}(\as^2)$. Parton distribution functions have
become increasingly more precise, largely due to the high-precision
HERA data. On the other hand, FFs, though increasing in
accuracy~\cite{pr:d62:054001,prl:85:5288,pr:d75:034018,np:b803:42,pr:d75:114010,pr:d76:074033,np:b734:50,epj:c61:603},
still lack the precision of the proton PDFs.

The data most widely used to extract the FFs comes from $\ele$
annihilations into charged
hadrons~\cite{prep:294:1,epj:c35:2004,pr:d41:2675,epj:c5:585,pl:b459:397,epj:c18:203,pl:b643:147,prep:399:71,pr:d37:1,zfp:c72:191,epj:c16:407,zfp:c47:187,pl:b311:408}.
These data are very precise and the predicted cross sections do not
depend on PDFs. However, they do not provide information on how to
disentangle quark and anti-quark contributions to the FFs and the
gluon fragmentation remains largely unconstrained. In addition, the $\ele$
data have poor statistics at large $z$, leading to large uncertainties
in this region of phase space. Several parameterisations of the FFs
exist~\cite{pr:d62:054001,prl:85:5288,pr:d75:034018}.

In the last few years, new one-particle inclusive measurements coming
from both $pp$
collisions~\cite{prl:98:252001,prl:91:241803,prl:97:152302,pr:c75:064901}
and DIS~\cite{Hillenbrand:phd:2005} became available. The inclusion of
these data in the extraction of the FFs yields a much more complete picture
of the fragmentation process and provides a direct handle on quark,
anti-quark and gluon contributions. A global QCD analysis of $\ele$,
$pp$ and DIS data is now available for several
hadrons~\cite{pr:d75:114010,pr:d76:074033}. This global FF set agrees
with the previous extractions, based on $\ele$ data alone, in the
regions of phase space which are also well constrained by $\ele$ data
alone.

\section{Experimental set-up}
A detailed description of the ZEUS detector can be found
elsewhere~\cite{pl:b293:465,zeus:1993:bluebook}. A brief outline of
the components most relevant for this analysis is given below.

Charged particles were tracked in the central tracking detector
(CTD)~\cite{nim:a279:290,npps:b32:181,nim:a338:254}, the microvertex
detector (MVD)~\cite{nim:a581:656} and the straw tube tracker
(STT)~\cite{nim:a535:191}. The CTD and MVD operated in a magnetic
field of $1.43$ T provided by a thin superconducting solenoid. The CTD
consisted of $72$~cylindrical drift-chamber layers, organised in nine
superlayers covering the polar-angle\footnote{The ZEUS coordinate system 
  is a right-handed Cartesian system, with the $Z$ axis pointing in
  the proton beam direction, referred to as the ``forward direction'',
  and the $X$ axis pointing towards the centre of HERA. The
  coordinate origin is at the nominal interaction point.} region 
\mbox{$15^\circ<\theta<164^\circ$}. 

The MVD silicon tracker consisted of a barrel (BMVD) and a forward
(FMVD) section. The BMVD contained three layers and provided
polar-angle coverage for tracks from $30^{\circ}$ to
$150^{\circ}$. The four-layer FMVD extended the polar-angle coverage
in the forward region to $7^{\circ}$. After alignment, the single-hit
resolution of the MVD was 24 $\mu$m. The transverse distance of
closest approach (DCA) to the nominal vertex in $X–Y$ was measured to
have a resolution, averaged over the azimuthal angle, of 
($46\oplus 122/p_T$) $\mu$m, with $p_T$ in GeV. The STT covered the
polar-angle region $5^\circ<\theta<25^\circ$. For CTD-MVD tracks
that pass through all nine CTD superlayers, the momentum resolution
was $\sigma(p_T)/p_T=0.0029p_T\oplus 0.0081\oplus 0.0012/p_T$, 
with $p_T$ in GeV.

The high-resolution uranium--scintillator calorimeter
(CAL)~\cite{nim:a309:77,nim:a309:101,nim:a321:356,nim:a336:23}
covered $99.7\%$ of the total solid angle and consisted of three parts:
the forward (FCAL), the barrel (BCAL) and the rear (RCAL) calorimeters. 
Each part was subdivided transversely into towers and longitudinally
into one electromagnetic section (EMC) and either one (in RCAL) or two
(in BCAL and FCAL) hadronic sections (HAC). The smallest subdivision
of the calorimeter was called a cell. Under test-beam conditions, the
CAL single-particle relative energy resolutions were
$\sigma(E)/E=0.18/\sqrt E$ for electrons and 
$\sigma(E)/E=0.35/\sqrt E$ for hadrons, with $E$ in GeV.

The energy of the scattered electron was corrected for energy loss
in the material between the interaction point and the calorimeter
using the small-angle rear tracking
detector~\cite{epj:c21:443,nim:a401:63} and the
presampler~\cite{epj:c21:443,nim:a382:419}.

The luminosity was measured using the Bethe-Heitler reaction 
$ep\rightarrow e\gamma p$ by the luminosity
detector~\cite{desy-92-066,zfp:c63:391,acpp:b32:2025} which
consisted of two independent systems. In the first system, the photons
were detected by a lead--scintillator calorimeter placed in the HERA
tunnel 107 m from the interaction point in the lepton-beam
direction. The second system was a magnetic spectrometer
arrangement~\cite{nim:a565:572}, which measured electron-positron
pairs from converted photons. The fractional uncertainty on the
measured luminosity was $1.8\%$.

\section{Event selection}
\label{eventsel}
The data used in this analysis were collected during the running
period 2005--2007, when HERA operated with protons of energy
$E_p=920$~GeV and electrons of energy $E_e=27.5$~GeV, and correspond
to an integrated luminosity of $330$~\pb1. The criteria to select DIS
events are described below.

A three-level trigger system~\cite{zeus:1993:bluebook,proc:chep:1992:222}
was used to select events online. It relied on the presence of an energy
deposition in the CAL compatible with that of a scattered electron.
At the third level, an identified electron~\cite{nim:a365:508} with
an energy larger than $4\gev$ was required.

Offline, the kinematic variables $\q2$, inelasticity, $y$, and the Bjorken
scaling variable, $x$, as well as the boost vector to the Breit frame
were reconstructed using the double-angle (DA)
method~\cite{proc:hera:1991:23}, which uses the angles of the
scattered electron and of the hadronic system.

Deep inelastic scattering events were selected by the following
requirements:
\begin{itemize}
\item $E_e^{\prime}>10$ GeV, where $E_e^{\prime}$ is the
  scattered-electron energy; this ensures a reconstruction
  efficiency above $95\%$ and a purity of the scattered electron of
  $\approx 100\%$;
\item $y_e\leq 0.95$, where $y_e$ is the inelasticity estimated from
  the energy and angle of the scattered electron; this excludes events
  with spurious electrons in the forward region, which are produced
  predominantly by photoproduction;
\item $y_{\rm JB}\geq 0.04$, where $y_{\rm JB}$ is the inelasticity
  estimated using the Jacquet-Blondel
  method \cite{proc:epfacility:1979:391}; this rejects events for
  which the DA method gives a poor reconstruction; 
\item $35<\delta<60$ GeV, where $\delta=\sum(E_i-P_{Z_i})$
  and $E_i$ is the energy of the $i$-th CAL cell, $P_{Z_i}$ is the
  momentum along the $Z$ axis and the sum runs over all CAL cells;
  this removes the phase space where photoproduction background and
  events with initial-state radiation are expected;
\item $|Z_{\rm vtx}|<50$~cm, where $Z_{\rm vtx}$ is the $Z$ component
  of the position of the primary vertex; this reduces
  background from events not originating from $ep$ collisions;
\item $|X|>12$ and $|Y|>12$ cm, where $X$ and $Y$ are the impact
  positions of the scattered electron on the RCAL, to avoid the
  low-acceptance region adjacent to the rear beampipe;
\item the analysis was restricted to events with $10<\q2<40000$~\g2\
  and $0.001<x<0.75$.
\end{itemize}

These requirements selected a sample of $2.16\cdot 10^7$ DIS data events.

\section{{\boldmath $\k0s$} and {\boldmath $\Lambda$} selection and
  reconstruction}
\label{klsel}
The strange hadrons $\k0s$ and $\Lambda$ were identified via the
charged-decay channels, $\k0s\rightarrow\pi^+\pi^-$ and 
$\Lambda\rightarrow p\pi^-$ ($\bar\Lambda\rightarrow\bar p\pi^+$).
The candidates were reconstructed using two
oppositely charged tracks associated with a displaced secondary
vertex. In the case of the $\k0s$, the mass of the pion was assigned
to both tracks. For the $\Lambda$, the mass of the proton was assigned
to the track with the largest momentum, whereas the mass of the pion was
assigned to the other track, since the proton always has a larger
momentum than the pion for $\Lambda$ baryons with momentum larger than
$0.3$~GeV.

All tracks were required to be in the region of high CTD acceptance, 
$|\eta^{\rm track}|<1.75$, where $\eta=-\ln(\tan\theta/2)$ is the
pseudorapidity in the laboratory frame and $\theta$ is the polar angle 
with respect to the proton beam direction. The tracks had to pass
through at least three CTD superlayers and were required to have
transverse momenta $P_T^{\rm track}>150$~MeV. 

The analysis was restricted to the current region of the Breit frame
by boosting the tracks to this frame and requiring $P_Z^{\rm
  Breit}<0$, where $P_Z^{\rm Breit}$ is the longitudinal momentum of
the track in the Breit frame. The combined four-vector momentum of
the two tracks in the Breit frame, $P^{\rm Breit}$, was used to
reconstruct $x_p$.

Additional selection criteria, similar to those used in a previous
analysis~\cite{pl:b652:1}, were applied to the selected candidates to
maximise the purity of the sample with a minimum loss of
statistics. These requirements were:

\begin{itemize}  
\item $dca<2$ cm, where $dca$ is the distance of
  closest approach of the two tracks forming the candidate;
\item $\chi^2/dof<5$ for the $\chi^2$ of the secondary vertex fit;
\item $M(\ele)>60$~MeV, to eliminate background from photon
  conversion;
\item $M(p\pi)>1121$~MeV ($M(\pi^+\pi^-)<475$~MeV), to eliminate
  $\Lambda$ ($\k0s$) background from the $\k0s$ ($\Lambda$) sample;
\item $\theta_{2D}<0.03$~rad, where $\theta_{2D}$ is the
  collinearity angle in the $XY$ plane between
  the $\k0s\ (\Lambda)$-candidate momentum vector and the vector
  defined by the interaction point and the $\k0s\ (\Lambda)$ decay vertex;
\item $\theta_{3D}<0.04$~rad, where $\theta_{3D}$ is the
  collinearity angle between the $\k0s\ (\Lambda)$-candidate
  three-momentum vector and the vector defined by the interaction
  point and the $\k0s\ (\Lambda)$ decay vertex;
\item $L_{XY}>0.5\ (1)$ cm, where $L_{XY}$ is the distance between the
   $\k0s\ (\Lambda)$-candidate decay vertex and the primary vertex in
   the transverse plane;
\item $P_T^{\rm PA}>(<)\ 0.11$ GeV, where $P_T^{\rm PA}$ is the
  projection of the pion momentum onto a plane perpendicular
  to the $\k0s$ ($\Lambda$) momentum direction (the Podolanski-Armenteros
  variable~\cite{phm:43:13}).
\end{itemize}

Figures~\ref{figk} and \ref{figl} show the $dca$, $\theta_{2D}$,
$\theta_{3D}$ and $L_{XY}$ distributions for data and Monte Carlo (see
Section~\ref{mc}) for $\k0s$ and $\Lambda$ candidates, respectively. The
description of the data by the Monte Carlo simulation is adequate.

Figure~\ref{fig1} shows the $M(\pi^+\pi^-)$ and $M(p\pi)$
distributions after these requirements. A small amount of background
is observed. The fit shown in Fig.~\ref{fig1} is for illustration only.
The number of $\k0s$ ($\Lambda$) candidates in each bin of $x_p$ and
$\q2$ was estimated by counting the entries in the signal region,
$472-522$ ($1107.0-1124.5$) MeV, and subtracting the number of expected
background entries. The latter was determined from a linear fit to the
sideband regions $403-422$ and $572-597$ ($1086.0-1098.2$ and
$1133.2-1144.4$) MeV, also indicated in Fig.~\ref{fig1}.
There were $806\; 505$ ($165\; 875$) $\k0s$ ($\Lambda$)
candidates in the data sample. In the current region of the Breit
frame, there were $238\; 153$ $\k0s$ and $40\; 728$ $\Lambda$
candidates. A Monte Carlo study showed that $6\%$ of the selected
$\Lambda$ candidates come from higher-baryon decays.

\section{Monte Carlo simulation}
\label{mc}
Samples of Monte Carlo (MC) events were produced to determine the
response of the detector and to correct the data to the hadron
level. The MC samples were also used to compute predictions to be
compared to the measurements.

The generated events were passed through the {\sc
  Geant}~3.21-based~\cite{tech:cern-dd-ee-84-1} ZEUS detector- and 
trigger-simulation programs~\cite{zeus:1993:bluebook}. They were
reconstructed and analysed by the same program chain as used for the
data. Particles with lifetime longer than $3\cdot 10^{-11}$~s, such
as $\k0s$ and $\Lambda$, were treated as stable at generator level and
their decays were simulated by {\sc Geant}.

Neutral current DIS events were generated using the program {\sc
  Lepto}~6.5.1~\cite{cpc:101:108}. Radiative effects were estimated
using the {\sc Heracles}~4.6.6~\cite{cpc:69:155,spi:www:heracles}
program with the {\sc Djangoh}~1.6~\cite{cpc:81:381,spi:www:djangoh11} 
interface to {\sc Lepto}. {\sc Heracles} includes QED corrections for
initial- and final-state radiation, vertex and propagator terms, and
two-boson exchange. The QCD cascade was simulated using the
colour-dipole
model (CDM)~\cite{pl:b165:147,pl:b175:453,np:b306:746,zfp:c43:625},
including the leading-order QCD diagrams as implemented in
{\sc Ariadne}~4.12~\cite{cpc:71:15,zfp:c65:285} and, alternatively,
with the MEPS model of {\sc Lepto}. Fragmentation into hadrons was
performed using the Lund string model~\cite{prep:97:31}, as implemented in 
{\sc Jetset}~7.41~\cite{cpc:82:74,cpc:135:238,cpc:39:347,cpc:43:367}. 
The default parameter setting from the
DELPHI/EMC~\cite{zfp:c35:417,pl:b311:408} tune was used for the
hadronisation. The CTEQ5D~\cite{epj:c12:375} proton PDFs were used for
these simulations.

\section{Corrections and systematic uncertainties}
\label{cor}
The measured scaled momentum distributions were corrected to the
hadron level and to the QED Born level. The correction factors were
calculated bin-by-bin using the MC samples described in
Section~\ref{mc}. The correction factors take into account: ({\em i})
the event-selection efficiency for the cuts listed in
Section~\ref{eventsel}, but for the $\q2$ and $x$ requirements; ({\em
  ii}) the efficiency to identify the $\k0s$ and $\Lambda$ decays, as
specified in Section~\ref{klsel}; ({\em iii}) the migrations between
bins due to detector resolution, which affects in particular the
transformation to the Breit frame; ({\em iv}) the relevant branching
ratios; and ({\em v}) the extrapolation to the full phase space. The
factors calculated in the measured $(x_p, \q2)$ bins varied from
$0.05\ (0.05)$ to $0.18\ (0.11)$ for $\k0s\ (\Lambda)$ candidates, and
reached $\approx 0.25$ for candidates with momentum in the range
$1-1.5$ GeV and $-1<\eta<1$; the lowest values were found for high
$\q2$ and $x_p$ values. Bins with an acceptance smaller than $0.05$
were not used in the analysis. The QED correction factors were
computed using the Monte Carlo samples; they are below $5\%$ for
$\q2<100$~\g2\ and increase to a maximum of $20\%$ at the highest
values of $\q2$.

The total systematic uncertainties on the scaled momentum distributions
are larger than the statistical uncertainties in most bins. The
statistical uncertainties themselves vary significantly over the
kinematic range. For $\k0s$ ($\Lambda$), they are at the $1\ (4)\%$
level at low $\q2$ and between $10$ to $90\%$ ($20$ to $70\%$) over
the $x_p$ range at large $\q2$. Many of the systematic
uncertainties were observed to scale with the statistical
uncertainty. In the following list, typical values of the
uncertainties on the scaled momentum distribution are given separately
for $\k0s$ and $\Lambda$, either as percentages of the statistical
uncertainty or as absolute values:

\begin{itemize}
\item imperfections in the simulation causing uncertainties on
  DIS event reconstruction and selection resulted in uncertainties of
  $^{+40}_{-30}\%$ and $^{+50}_{-40}\%$ of the statistical uncertainties.
  This was evaluated by modifying the selection cuts within the
  experimental resolutions. At low $\q2$, the variation of the cut 
  on $y_{\rm JB}$ from $0.04$ to $0.07$ resulted in large
  uncertainties exceeding these typical values;
\item an uncertainty of $-2\%$ in the overall tracking efficiency
  resulted in absolute uncertainties of $+4\%$ and $+4\%$;
\item detector-alignment uncertainties affecting the calculation of
  the boost vector to the Breit frame resulted in uncertainties of
  $^{+30}_{-25}\%$ and $^{+20}_{-15}\%$ of the statistical uncertainties.
  This was evaluated by varying separately the simulated polar angle of the
  scattered electron and of the hadrons by $\pm 2$~mrad;  
\item uncertainties on the $\k0s$ and $\Lambda$ selection
  efficiency resulted in uncertainties of $^{+80}_{-60}\%$ and
  $^{+60}_{-60}\%$ of the statistical uncertainties. This was
  evaluated by varying the cuts listed in Section~\ref{klsel}: the
  dominant effects were due to modifications of the cuts on $\theta_{2D}$ to
  0.015 and 0.06 and $\theta_{3D}$ to 0.02 and 0.08;
\item assumptions concerning the details of the simulation of the
  hadronic final state resulted in absolute uncertainties of
  $^{+4}_{-3}\%$ and $^{+10}_{-15}\%$. At large $\q2$, these
  uncertainties were larger and exceeded $^{+15}_{-80}\%$ and
  $^{+50}_{-25}\%$. This was estimated by using MEPS instead of
  CDM in the calculation of the correction factors;
\item background-subtraction uncertainties resulted in absolute
  uncertainties of $^{+2}_{-2}\%$ and $^{+3}_{-4}\%$. At large $\q2$, the
  uncertainties exceeded these typical values and were as high as $\pm
  35\%$ for both $\k0s$ and $\Lambda$. This was evaluated by varying
  the size of the background window by $\pm 40\%$ and changing the
  background fit function from first to second order.
\end{itemize} 

The systematic uncertainties were added in quadrature for each
bin. The total systematic uncertainty is dominated by the uncertainty
in the simulation of the hadronic final state. At low $\q2$, the
overall tracking efficiency also contributes significantly. At high
$\q2$, the uncertainties related to the $\k0s$ and $\Lambda$ selection
are important.

\section{NLO QCD calculations}
\label{nlo}
Next-to-leading-order QCD calculations, which combine the full NLO
matrix elements with the proton PDFs and FFs as explained in
Section~\ref{theory}, were compared to the measurements. For the
comparison, the observable $x_p$ is assumed to be equal to the
variable $z$. For each bin in $x_p$ and $\q2$, a prediction was
derived by numerical integration over the multiplicities
$d^2m(H)/dzd\q2$, with $m(H)$ the number of $H$ per DIS event. Two
sets of calculations based on different parameterisations of the FFs
were used. The first set was obtained from fits to $\ele$ data and
based on the program {\sc Cyclops}~\cite{pl:b406:178}, called ``AKK+{\sc
  Cyclops}''~\cite{pr:d75:034018,np:b803:42}. The second set was
obtained from a global fit to $\ele$, $pp$ and $ep$ data, called
``DSS''~\cite{pr:d75:114010}. It was used only for $\k0s$ predictions.

The AKK+{\sc Cyclops} calculations were performed using $Q$ as the
factorisation and renormalisation scales; the number of active quark
flavours was set to $n_f=5$; the proton PDFs were parameterised
using the CTEQ6M sets~\cite{jhep:0207:012} and $\Lambda_{\rm QCD}$ was
set to 226~MeV. The calculations were done assuming massless particles.
Hadron-mass effects~\cite{prl:95:232002} for $\k0s$ and $\Lambda$ were
included as correction factors~\cite{np:b803:42}. The influence on the
shapes of the calculated scaled momentum distributions due to the mass
effects is expected at small values of $x_p$ and $\q2$, as explained
in Section~\ref{theory}.

In the DSS calculations, the scaled momentum distributions were
obtained by convoluting the NLO DSS set of FFs together with the MRST
NLO~\cite{pl:b531:216} PDFs and appropriate NLO coefficient
functions. For these calculations, $\k0s$-mass corrections were not
included. The predictions were computed as ratios for each bin, such
that a later combination of bins is not possible~\cite{pc}.

The uncertainty from terms beyond NLO was estimated by varying the
renormalisation scale by factors 0.5 and 2. The uncertainties from FFs
could not be evaluated so far; it is to a certain extent represented
by the differences in the predictions of AKK+CYCLOPS and DSS. In
addition, it should be noted that the DSS FFs were extracted from
data at low $\q2$ and that the fits are thus almost unconstrained at
high $\q2$~\cite{pr:d75:114010}.

\section{Results}
\label{results}
Scaled momentum distributions, $(1/N)(n(H)/\Delta x_p)$, with $n(H)$
the number of $H$ ($\k0s$ or $\Lambda$), $N$ the number of DIS
events in a given $\q2$ bin and $\Delta x_p$ the width of the $x_p$ bin,
were measured in the current region of the Breit frame. The
distributions are presented as functions of $\q2$ and $x_p$ in
the kinematic region of $10<\q2<40000$~\g2\ and $0.001<x<0.75$.

Figure~\ref{fig2} shows the scaled momentum distributions for $\k0s$ as 
functions of $\q2$ in different regions of $x_p$. The results are also
presented in Table~\ref{tab1}. The data show clear scaling
violation. This behaviour is expected on the basis of the QCD
description of the parton evolution with increasing $Q$: the phase
space for soft gluon radiation increases, leading to a rise of the
number of soft particles with small $x_p$.

The predictions from the CDM and MEPS models, based on
leading-logarithmic matrix elements plus parton shower and the Lund 
fragmentation model, as described in Section~\ref{mc}, are compared to
the measurements in Fig.~\ref{fig2}. They describe the shapes of the
distributions fairly well while overestimating the overall production
of $\k0s$ by $10$ to $20\%$.

The NLO QCD calculations, based on full NLO matrix elements and the
fragmentation-function approach described in Sections~\ref{theory} and
\ref{nlo}, are also compared to the measurements in Fig.~\ref{fig2} for
$x_p>0.1$. For $z<0.1$, the calculations become singular.

The AKK+{\sc Cyclops} calculations, based on FFs extracted from $\ele$
data alone, fail to describe the measurements. These calculations
predict a much too high $\k0s$ rate but for $x_p>0.6$. These
discrepancies might come from the fact that the FFs used in these
predictions have a poorly constrained gluon contribution, which is
dominant at low $x_p$.

The DSS calculations, based on FFs extracted from a global analysis,
give a good description of the measurements for $x_p>0.3$ and
$10<\q2<40000$~\g2. The prediction for this region of phase space
is mainly constrained by $pp$ data, which sufficiently constrain the
FFs at high $x_p$. At lower $x_p$, the DSS calculations fail to
describe the data. This can be explained 
by the fact that the DSS fit in this region of phase space is mostly
unconstrained by the available data. Thus, the measurements presented
in this paper will help to improve significantly such global fits in
this region of phase space.

Figure~\ref{fig3} and Table~\ref{tab3} show the scaled momentum
distributions for $\k0s$ as functions of $x_p$ in two regions of
$\q2$. The predictions of CDM and MEPS give a good description of the
data. In both regions of $\q2$, both NLO calculations predict
too-steep spectra. At low $\q2$, this effect is especially pronounced.

Figures~\ref{fig4} and \ref{fig5} show the scaled momentum
distributions for $\Lambda$. The results are also presented in
Tables~\ref{tab4} and \ref{tab6}. Scaling violations are clearly
observed. The predictions of CDM and MEPS give a reasonable
description of the measurements, but overestimate the overall
$\Lambda$ rate by $\approx 20\%$. The AKK+{\sc Cyclops} NLO
calculations fail to describe the measurements. As seen in
Fig.~\ref{fig5}, the predicted spectra in $x_p$ are, as in the case of
$\k0s$, significantly too steep.

ZEUS has previously published measurements of scaled momentum
distributions for inclusive charged particles in
DIS~\cite{jhep:1006:009}. These measurements are dominated by the
contribution from charged pions. Figure~\ref{fig6} shows 
the scaled momentum distributions presented in this paper together
with those from the inclusive charged particles analysis in the kinematic
region of $0.1<x_p<0.4$ as functions of $\q2$. For $\q2>100$~\g2, all
distributions show a plateau. At lower $\q2$, and especially at low
$x_p$, sizeable mass effects are expected. This is clearly
visible. For $0.1<x_p<0.2$, the value of $(1/N)(n(H)/\Delta x_p)$
drops to $10\ (20)\%$ of its maximum value for $\Lambda\ (\k0s)$,
while for inclusive charged particles, the $(1/N)(n(H)/\Delta x_p)$
value is still $40\%$ of the plateau value at the lowest $\q2$
accessible.

\section{Summary and conclusions}
Scaled momentum distributions for $\k0s$ and $\Lambda$ hadrons
were measured for the first time in $ep$ DIS. The distributions were
measured in the $\q2$ range from 10 to 40000~\g2\ and
$0.001<x<0.75$. Scaling violations were clearly observed for both the
$\k0s$ and $\Lambda$ hadrons.

Next-to-leading-order QCD calculations, based on different
parameterisations of the FFs, were compared to the measurements. The
predictions based on FFs extracted from $\ele$ data alone fail to
describe the measurements. Those predictions based on a global analysis
which include $\ele$, $pp$ and $ep$ data give an improved description
of the measurements. However, they predict a too high production rate
of $\k0s$ and $\Lambda$ hadrons at low $x_p$ and $\q2$. The
measurements presented in this paper have the potential to constrain
significantly the FFs for the strange hadrons $\k0s$ and $\Lambda$.

\vspace{0.5cm}
\noindent {\Large\bf Acknowledgements}
\vspace{0.3cm}

We thank the DESY Directorate for their strong support and
encouragement. The remarkable achievements of the HERA machine group
were essential for the successful completion of this work and are
greatly appreciated. We are grateful for the support of the DESY
computing and network services. The design, construction and
installation of the ZEUS detector have been made possible owing to the
ingenuity and effort of many people who are not listed as authors. 
We would like to thank S. Albino, R. Sassot and collaborators
for providing their calculations. Special thanks are due to R. Sassot
for very useful discussions.

\newpage
\providecommand{\etal}{et al.\xspace}
\providecommand{\coll}{Collaboration}
\catcode`\@=11
\def\@bibitem#1{%
\ifmc@bstsupport
  \mc@iftail{#1}%
    {;\newline\ignorespaces}%
    {\ifmc@first\else.\fi\orig@bibitem{#1}}
  \mc@firstfalse
\else
  \mc@iftail{#1}%
    {\ignorespaces}%
    {\orig@bibitem{#1}}%
\fi}%
\catcode`\@=12

\newpage
\clearpage
\begin{table}
\begin{center}
\begin{tabular}{|c||c|c|c|} \hline
 & & &    \\
 $\q2$ (\g2) & 0.0 $< x_p <$ 0.1 & 0.1 $< x_p <$ 0.2 & 0.2 $< x_p <$ 0.3  \\ 
 & & &   \\
\hline
\hline
 & & &    \\
10 - 40 & 0.031$\pm0.001^{+0.002}_{-0.001}$ & 0.125$\pm0.001^{+0.008}_{-0.003}$ & 0.144$\pm0.001^{+0.009}_{-0.013}$  \\
 & & &   \\
\hline
 & & &    \\
40 - 160 &  0.171$\pm0.002^{+0.009}_{-0.006}$ & 0.392$\pm0.003^{+0.018}_{-0.010}$ & 0.283$\pm0.002^{+0.013}_{-0.006}$  \\
 & & &    \\
\hline
 & & &    \\
160 - 640 &  0.551$\pm0.010^{+0.025}_{-0.018}$ & 0.612$\pm0.010^{+0.028}_{-0.017}$ & 0.306$\pm0.007^{+0.014}_{-0.009}$  \\
 & & &   \\
\hline
 & & &    \\
640 - 2560 & 1.141$\pm0.038^{+0.087}_{-0.037}$ & 0.618$\pm0.030^{+0.1130}_{-0.016}$ & 0.309$\pm0.029^{+0.047}_{-0.011}$ \\
 & & &   \\
\hline
 & & &   \\
2560 - 10240 &  1.878$\pm0.168^{+0.095}_{-0.147}$ & 0.834$\pm0.217^{+0.065}_{-0.278}$ & 0.115$\pm0.062^{+0.066}_{-0.054}$ \\
 & & &    \\
\hline
\hline
 & & &   \\
 $\q2$ (\g2) & 0.3 $< x_p <$ 0.4 & 0.4 $< x_p <$ 0.6 & 0.6 $< x_p <$ 1.0  \\ 
 & & &  \\
\hline
\hline
 & & &   \\
10 - 40 & 0.1112$\pm0.0008^{+0.0074}_{-0.0130}$ & 0.0130$\pm0.0004^{+0.0041}_{-0.0013}$ & 0.0132$\pm0.0001^{+0.0007}_{-0.0004}$  \\
 & & &   \\
\hline
 & & &   \\
40 - 160 & 0.1571$\pm0.0019^{+0.0082}_{-0.0033}$ &  0.0671$\pm0.0009^{+0.0038}_{-0.0014}$ & 0.0109$\pm0.0003^{+0.0006}_{-0.0002}$  \\
 & & &   \\
\hline
 & & &   \\
160 - 640 & 0.1585$\pm0.0060^{+0.0118}_{-0.0051}$ & 0.0548$\pm0.0027^{+0.0050}_{-0.0014}$ & 0.0073$\pm0.0008^{+0.0005}_{-0.0011}$   \\
 & & &   \\
\hline
 & & &    \\
640 - 2560 & 0.1053$\pm0.0217^{+0.0319}_{-0.0214}$ & 0.0558$\pm0.0141^{+0.0176}_{-0.0028}$ & 0.0029$\pm0.0022^{+0.0014}_{-0.0030}$  \\
 & & &   \\
\hline
\end{tabular}
\end{center}
\caption
{\it The measured scaled momentum distributions $(1/N)(n(\k0s)/\Delta x_p)$
as functions of $\q2$ in different regions of $x_p$. The statistical
and systematic uncertainties are also shown.}
\label{tab1}
\end{table}

\newpage
\clearpage
\begin{table}
\begin{center}
\begin{tabular}{|c||c|c|} \hline
 & &    \\
 $x_p$ & $10<\q2<100$~\g2 & $100<\q2<40000$~\g2  \\ 
 & &   \\
\hline
\hline
 & &    \\
0.0 - 0.1 & 0.0488$\pm0.0006^{+0.0024}_{-0.0012}$ & 0.4841$\pm0.0063^{+0.0233}_{-0.0117}$   \\
 & &    \\
\hline
 & &    \\
0.1 - 0.2 &  0.1618$\pm0.0009^{+0.0094}_{-0.0041}$ & 0.5740$\pm0.0061^{+0.0263}_{-0.0132}$  \\
 & &    \\
\hline
 & &    \\
0.2 - 0.3 &  0.1648$\pm0.0009^{+0.0098}_{-0.0035}$ & 0.3140$\pm0.0048^{+0.0142}_{-0.0082}$  \\
 & &    \\
\hline
 & &    \\
0.3 - 0.4 & 0.1183$\pm0.0007^{+0.0077}_{-0.0026}$ & 0.1588$\pm0.0037^{+0.01213}_{-0.0045}$  \\
 & &    \\
\hline
 & &    \\
0.4 - 0.5 &  0.0751$\pm0.0006^{+0.0055}_{-0.0015}$ & 0.0760$\pm0.0027^{+0.0073}_{-0.0017}$  \\
 & &    \\
\hline
 & &    \\
0.5 - 0.6 & 0.0452$\pm0.0004^{+0.0031}_{-0.0010}$ & 0.0408$\pm0.0022^{+0.0037}_{-0.0011}$   \\
 & &    \\
\hline
 & &    \\
0.6 - 0.7 &  0.0260$\pm0.0003^{+0.0017}_{-0.0006}$ & 0.0182$\pm0.0015^{+0.0011}_{-0.0009}$  \\
 & &    \\
\hline
 & &    \\
0.7 - 0.8 &  0.0150$\pm0.0002^{+0.0009}_{-0.0003}$ & 0.0101$\pm0.0012^{+0.0008}_{-0.0016}$  \\
 & &    \\
\hline
 & &    \\
0.8 - 0.9 & 0.0073$\pm0.0001^{+0.0005}_{-0.0003}$ & 0.0034$\pm0.0007^{+0.0007}_{-0.0001}$  \\
 & &    \\
\hline
 & &    \\
0.9 - 1.0 &  0.0032$\pm0.0001^{+0.0002}_{-0.0006}$ & 0.0020$\pm0.0006^{+0.0008}_{-0.0001}$  \\
 & &    \\
\hline
\end{tabular}
\end{center}
\caption
{\it 
The measured scaled momentum distributions $(1/N)(n(\k0s)/\Delta x_p)$
as functions of $x_p$ in different regions of $\q2$. Other details as
in the caption to Table~\ref{tab1}.
}
\label{tab3}
\end{table}

\newpage
\clearpage
\begin{table}
\begin{center}
\begin{tabular}{|c||c|c|c|} \hline
 & & &   \\
 $\q2$ (\g2) & 0.0 $< x_p <$ 0.1 & 0.1 $< x_p <$ 0.2 & 0.2 $< x_p <$ 0.3  \\ 
 & & &  \\
\hline
\hline
 & & &   \\
10 - 40 & 0.0025$\pm0.0002^{+0.0003}_{-0.0002}$ & 0.0122$\pm0.0004^{+0.0010}_{-0.0008}$ & 0.0189$\pm0.0005^{+0.0015}_{-0.0014}$  \\
 & & &   \\
\hline
 & & &   \\
40 - 160 &  0.0189$\pm0.0014^{+0.0012}_{-0.0007}$ & 0.0650$\pm0.0021^{+0.0030}_{-0.0039}$ & 0.0656$\pm0.0018^{+0.0037}_{-0.0036}$  \\
 & & &   \\
\hline
 & & &   \\
160 - 640 &  0.0995$\pm0.0095^{+0.0085}_{-0.0073}$ & 0.1666$\pm0.0099^{+0.0091}_{-0.0163}$ & 0.0960$\pm0.0066^{+0.0074}_{-0.0038}$  \\
 & & &   \\
\hline
 & & &   \\
640 - 2560& 0.2313$\pm0.0427^{+0.0593}_{-0.0567}$ & 0.1966$\pm0.0414^{+0.0124}_{-0.0303}$ & 0.1038$\pm0.0346^{+0.0132}_{-0.0090}$ \\
 & & &   \\
\hline
 & & &   \\
2560 - 10240 &  0.7416$\pm0.4386^{+0.0577}_{-0.3574}$ & 0.1962$\pm0.1268^{+0.0393}_{-0.1395}$ &  \\
 & & &   \\
\hline
\hline
 & & &   \\
 $\q2$ (\g2) & 0.3 $< x_p <$ 0.4 & 0.4 $< x_p <$ 0.6 & 0.6 $< x_p <$ 1.0  \\ 
 & & &  \\
\hline
\hline
 & & &   \\
10 - 40 & 0.0184$\pm0.0004^{+0.0011}_{-0.0013}$ & 0.0106$\pm0.0002^{+0.0006}_{-0.0004}$ & 0.0021$\pm0.0001^{+0.0001}_{-0.0001}$  \\
 & & &   \\
\hline
 & & &   \\
40 - 160 & 0.0453$\pm0.0015^{+0.0028}_{-0.0031}$ &  0.0198$\pm0.0007^{+0.0011}_{-0.0010}$ & 0.0020$\pm0.0002^{+0.0003}_{-0.0001}$  \\
 & & &   \\
\hline
 & & &   \\
160 - 640 & 0.0641$\pm0.0074^{+0.0034}_{-0.0149}$ & 0.0139$\pm0.0025^{+0.0089}_{-0.0031}$ & 0.0041$\pm0.0009^{+0.0003}_{-0.0008}$   \\
 & & &   \\
\hline
 & & &   \\
640 - 2560 & 0.0653$\pm0.0436^{+0.0119}_{-0.0244}$ &  &  \\
 & & &   \\
\hline
\end{tabular}
\end{center}
\caption
{\it 
The measured scaled momentum distributions $(1/N)(n(\Lambda)/\Delta x_p)$
as functions of $\q2$ in different regions of $x_p$. Other details as
in the caption to Table.~\ref{tab1}.
}
\label{tab4}
\end{table}

\newpage
\clearpage
\begin{table}
\begin{center}
\begin{tabular}{|c||c|c|} \hline
 & &    \\
 $x_p$ & $10<\q2<100$~\g2 & $100<\q2<40000$~\g2  \\ 
 & &   \\
\hline
\hline
 & &    \\
0.0 - 0.1 & 0.00437$\pm0.00034^{+0.00046}_{-0.00017}$ & 0.08307$\pm0.00638^{+0.00626}_{-0.00826}$   \\
 & &    \\
\hline
 & &    \\
0.1 - 0.2 &  0.01822$\pm0.00059^{+0.00106}_{-0.00109}$ & 0.13904$\pm0.00466^{+0.00677}_{-0.01515}$  \\
 & &    \\
\hline
 & &    \\
0.2 - 0.3 &  0.02456$\pm0.00059^{+0.00163}_{-0.00118}$ & 0.09489$\pm0.00275^{+0.00541}_{-0.00646}$  \\
 & &    \\
\hline
 & &    \\
0.3 - 0.4 & 0.02173$\pm0.00049^{+0.00109}_{-0.00138}$ & 0.056013$\pm0.00188^{+0.00323}_{-0.00353}$  \\
 & &    \\
\hline
 & &    \\
0.4 - 0.5 &  0.01387$\pm0.00035^{+0.00088}_{-0.00030}$ & 0.02950$\pm0.00135^{+0.00181}_{-0.00310}$  \\
 & &    \\
\hline
 & &    \\
0.5 - 0.6 & 0.00913$\pm0.00026^{+0.00054}_{-0.00043}$ & 0.014640$\pm0.00095^{+0.00091}_{-0.00048}$   \\
 & &    \\
\hline
 & &    \\
0.6 - 0.7 &  0.00483$\pm0.00018^{+0.00028}_{-0.00027}$ & 0.00534$\pm0.00051^{+0.00068}_{-0.00034}$  \\
 & &    \\
\hline
 & &    \\
0.7 - 0.8 &  0.00245$\pm0.00011^{+0.00015}_{-0.00024}$ & 0.00178$\pm0.00028^{+0.00032}_{-0.00049}$  \\
 & &    \\
\hline
 & &    \\
0.8 - 0.9 & 0.00096$\pm0.00006^{+0.0011}_{-0.00009}$ & 0.00056$\pm0.00019^{+0.00006}_{-0.00018}$  \\
 & &    \\
\hline
 & &    \\
0.9 - 1.0 &  0.00038$\pm0.00004^{+0.00003}_{-0.00013}$ & 0.00032$\pm0.00013^{+0.00009}_{-0.00023}$  \\
 & &    \\
\hline
\end{tabular}
\end{center}
\caption
{\it 
The measured scaled momentum distributions $(1/N)(n(\Lambda)/\Delta x_p)$
as functions of $x_p$ in different regions of $\q2$. Other details as
in the caption to Table.~\ref{tab1}.
}
\label{tab6}
\end{table}

\newpage
\clearpage
\begin{figure}[p]
\vfill
\setlength{\unitlength}{1.0cm}
\begin{picture} (18.0,15.0)
\put (0.0,0.0){\centerline{\epsfig{figure=\figdir 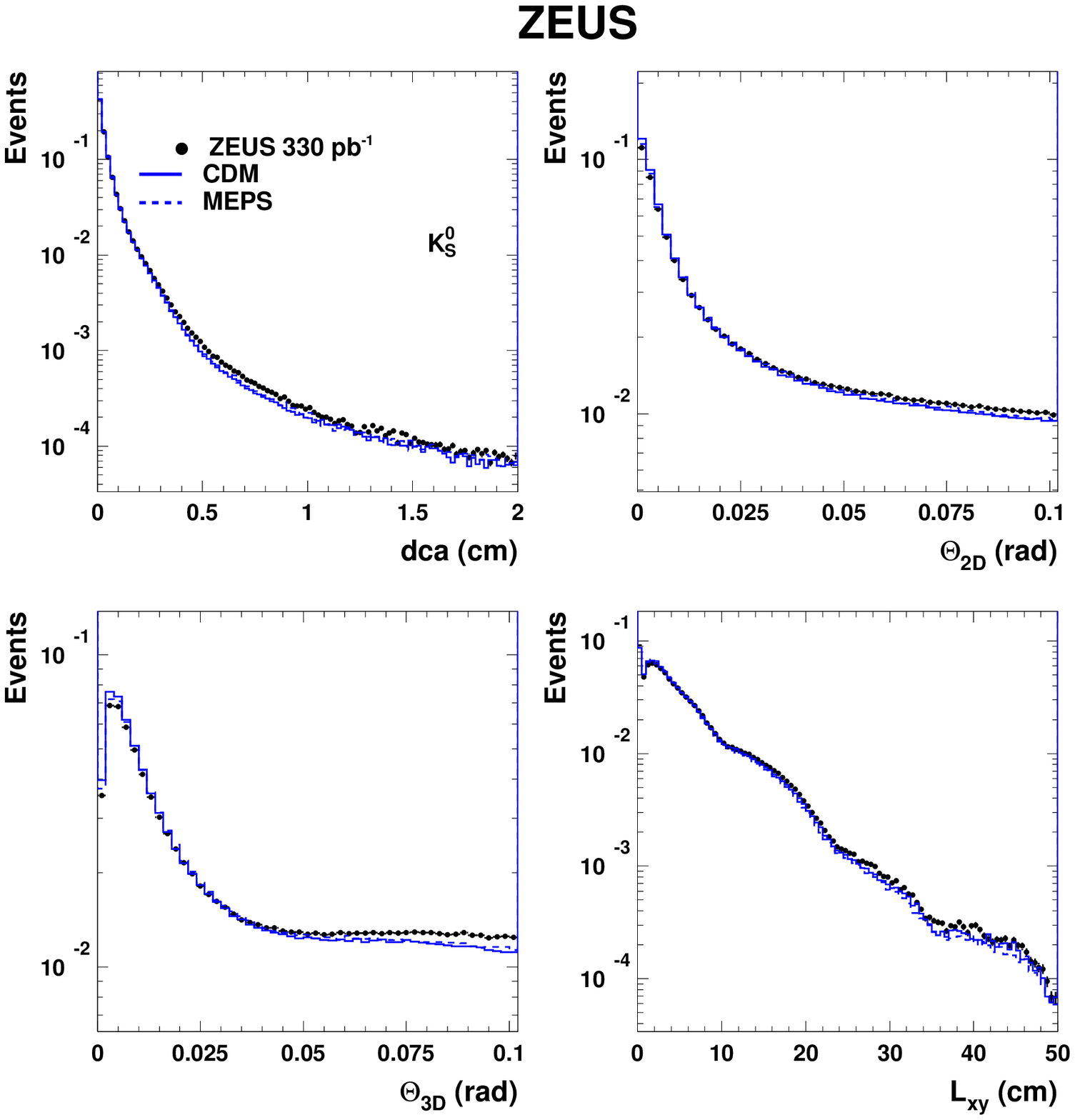,width=15cm}}}
\put (6.0,12.5){{\bf (a)}}
\put (13.0,12.5){{\bf (b)}}
\put (6.0,6.0){{\bf (c)}}
\put (13.0,6.0){{\bf (d)}}
\end{picture}
\caption
{\it 
The normalised (a) $dca$, (b) $\theta_{2D}$, (c) $\theta_{3D}$ and (d)
$L_{XY}$ distributions for data (dots) and Monte Carlo (histograms)
for $\k0s$ candidates.
}
\label{figk}
\vfill
\end{figure}

\newpage
\clearpage
\begin{figure}[p]
\vfill
\setlength{\unitlength}{1.0cm}
\begin{picture} (18.0,15.0)
\put (0.0,0.0){\centerline{\epsfig{figure=\figdir 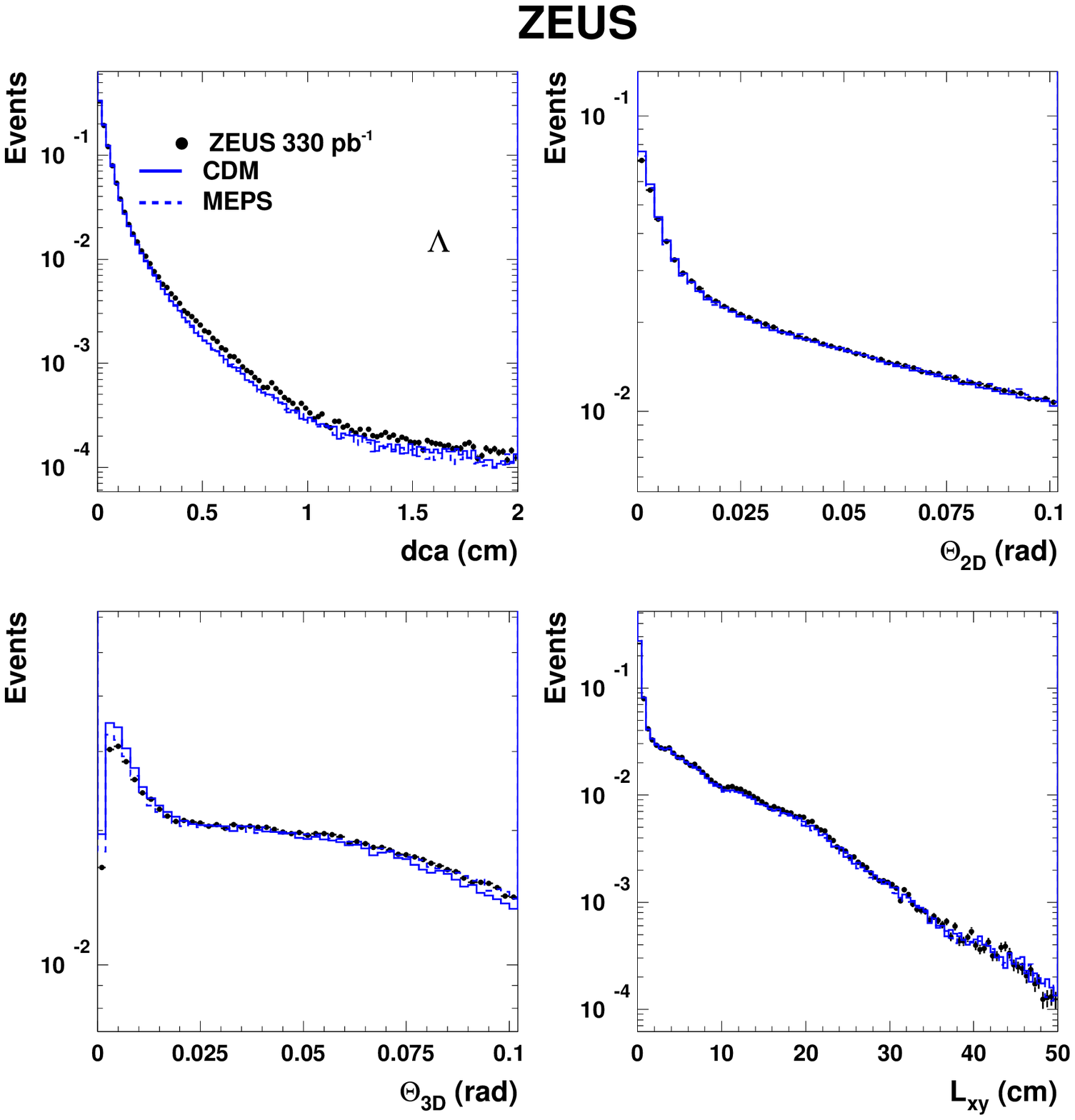,width=15cm}}}
\put (6.0,12.5){{\bf (a)}}
\put (13.0,12.5){{\bf (b)}}
\put (6.0,6.0){{\bf (c)}}
\put (13.0,6.0){{\bf (d)}}
\end{picture}
\caption
{\it 
The normalised (a) $dca$, (b) $\theta_{2D}$, (c) $\theta_{3D}$ and (d)
$L_{XY}$ distributions for data (dots) and Monte Carlo (histograms)
for $\Lambda$ candidates.
}
\label{figl}
\vfill
\end{figure}

\newpage
\clearpage
\begin{figure}[p]
\vfill
\setlength{\unitlength}{1.0cm}
\begin{picture} (18.0,15.0)
\put (0.0,0.0){\centerline{\epsfig{figure=\figdir 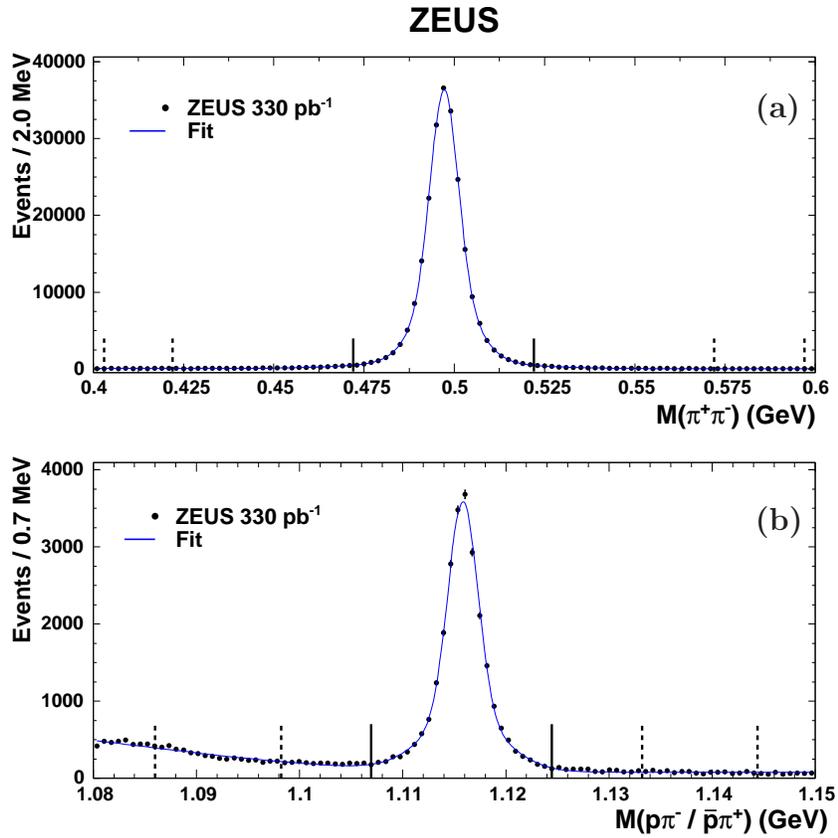,width=12cm}}}
\put (12.0,10.0){{\bf (a)}}
\put (12.0,4.5){{\bf (b)}}
\end{picture}
\caption
{\it 
(a) The $\pi^+\pi^-$ invariant-mass distribution for
$\k0s$ candidates (dots). (b) The $p\pi^-/\bar p\pi^+$ invariant-mass
distribution for $\Lambda/\bar\Lambda$ candidates
(dots). In both (a) and (b), the solid line represents an indicative fit
by two Gaussians and a (a) linear and (b) quadratic background function.
The solid vertical lines indicate the signal window used in the
analysis. The dashed lines indicate the two sideband regions used for
the background subtraction in each kinematic bin.
}
\label{fig1}
\vfill
\end{figure}

\newpage
\clearpage
\begin{figure}[p]
\vfill
\setlength{\unitlength}{1.0cm}
\begin{picture} (18.0,18.0)
\put (0.0,0.0){\centerline{\epsfig{figure=\figdir 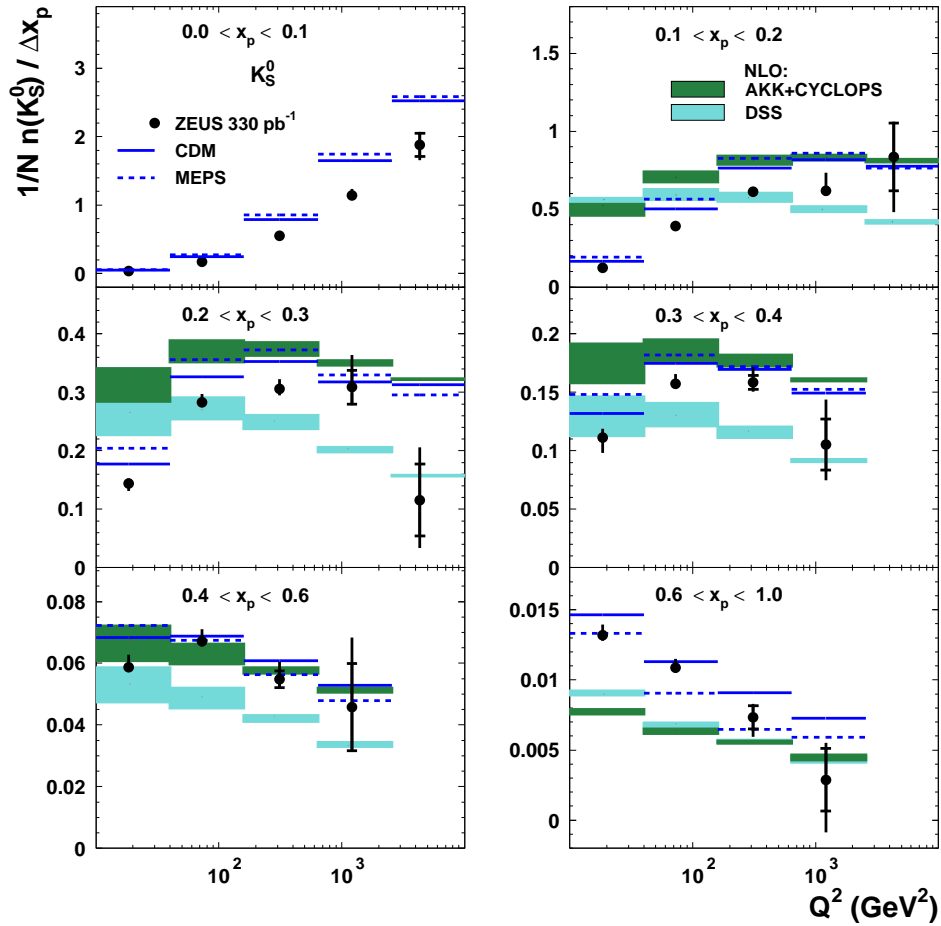,width=14cm}}}
\end{picture}
\caption
{\it 
The measured scaled momentum distributions $(1/N)(n(\k0s)/\Delta x_p)$
as functions of $\q2$ in different regions of $x_p$ (dots). The inner
error bars represent the statistical uncertainty. The outer error bars
show the statistical and systematic uncertainties added in
quadrature. In some bins, the error bars on the data points are
smaller than the marker size and are therefore not visible. For
comparison, the NLO predictions of AKK+{\sc Cyclops} (dark-shaded
band) and DSS (light-shaded band) are also presented. The bands
represent the theoretical uncertainty. The predictions from CDM (solid
lines) and MEPS (dashed lines) are also shown.
}
\label{fig2}
\vfill
\end{figure}

\newpage
\clearpage
\begin{figure}[p]
\vfill
\setlength{\unitlength}{1.0cm}
\begin{picture} (18.0,18.0)
\put (0.0,0.0){\centerline{\epsfig{figure=\figdir 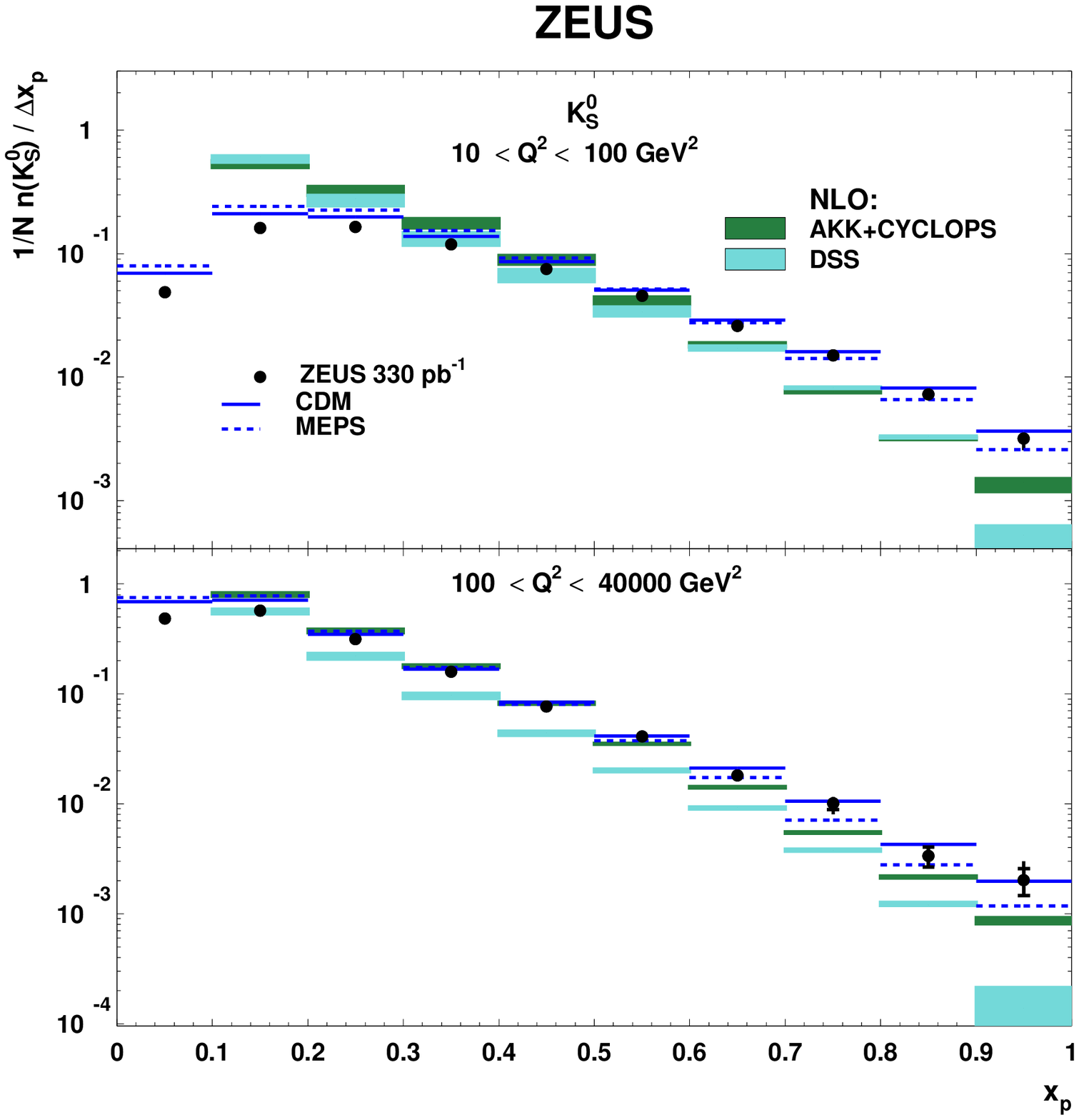,width=14cm}}}
\end{picture}
\caption
{\it 
The measured scaled momentum distributions $(1/N)(n(\k0s)/\Delta x_p)$
as functions of $x_p$ in different regions of $\q2$ (dots). Other
details as in the caption to Fig.~\ref{fig2}.
}
\label{fig3}
\vfill
\end{figure}

\newpage
\clearpage
\begin{figure}[p]
\vfill
\setlength{\unitlength}{1.0cm}
\begin{picture} (18.0,18.0)
\put (0.0,0.0){\centerline{\epsfig{figure=\figdir 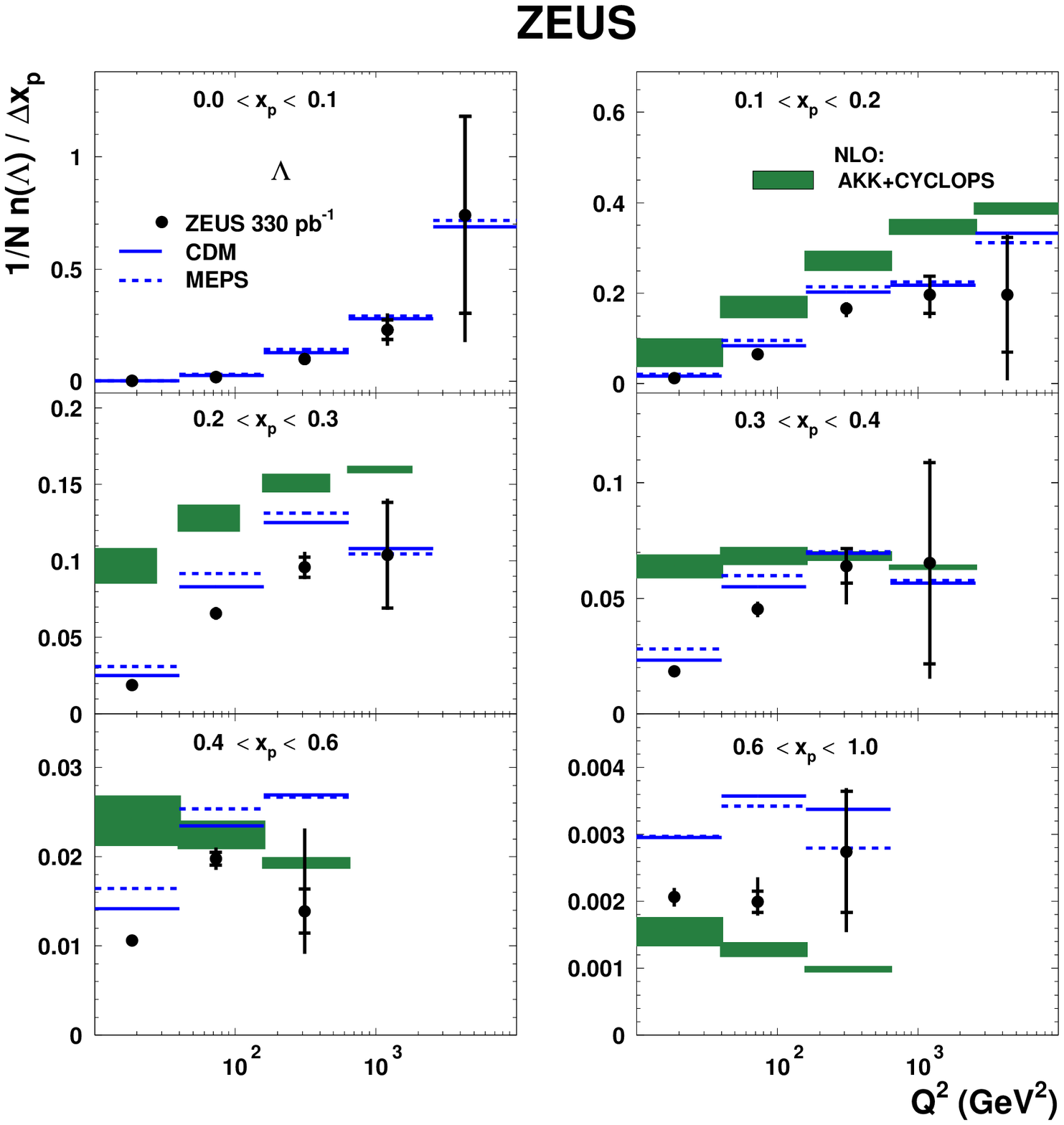,width=14cm}}}
\end{picture}
\caption
{\it 
The measured scaled momentum distributions $(1/N)(n(\Lambda)/\Delta x_p)$
as functions of $\q2$ in different regions of $x_p$ (dots). Other
details as in the caption to Fig.~\ref{fig2}.
}
\label{fig4}
\vfill
\end{figure}

\newpage
\clearpage
\begin{figure}[p]
\vfill
\setlength{\unitlength}{1.0cm}
\begin{picture} (18.0,18.0)
\put (0.0,0.0){\centerline{\epsfig{figure=\figdir 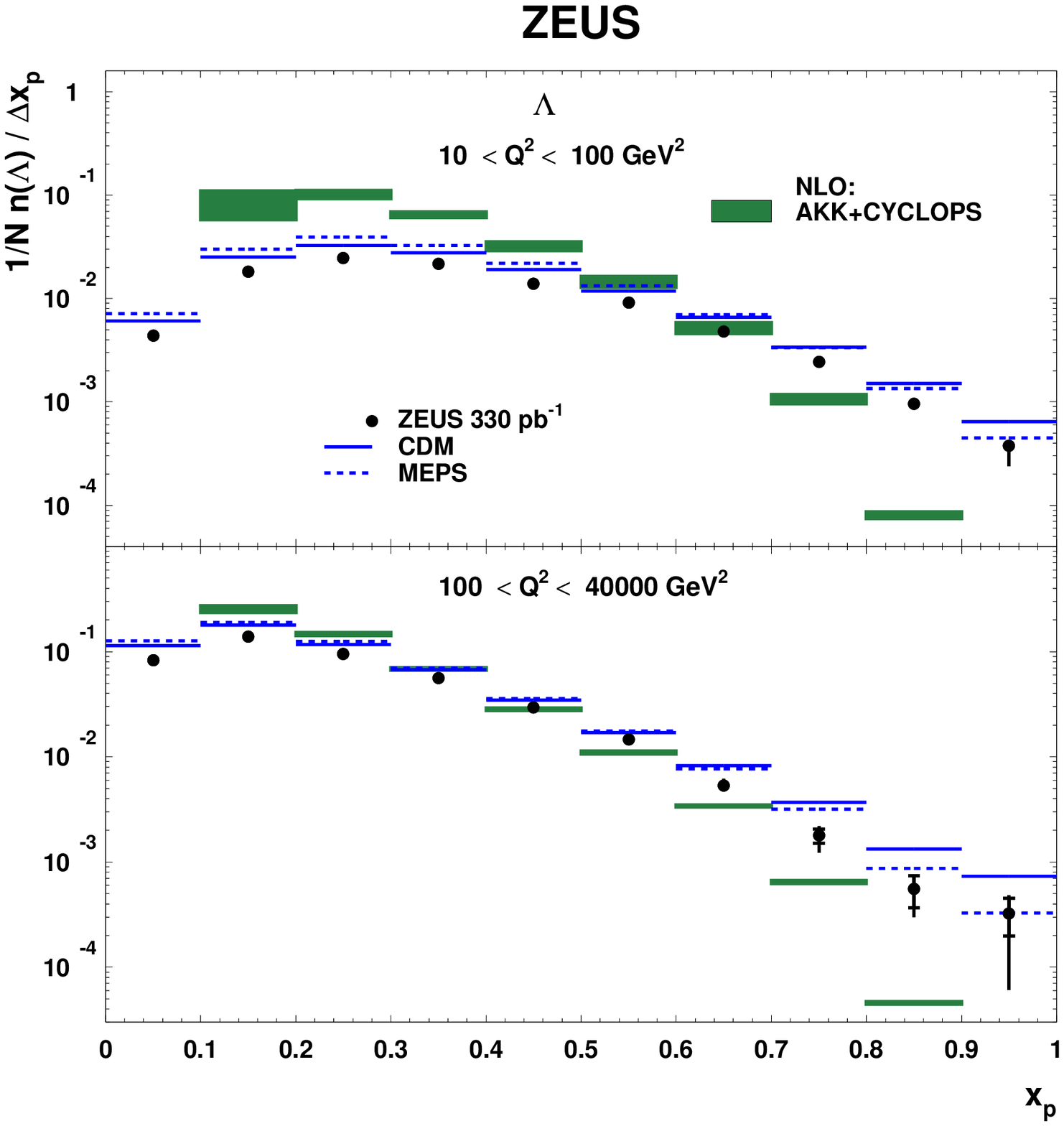,width=14cm}}}
\end{picture}
\caption
{\it 
The measured scaled momentum distributions $(1/N)(n(\Lambda)/\Delta x_p)$
as functions of $x_p$ in different regions of $\q2$ (dots). Other
details as in the caption to Fig.~\ref{fig2}.
}
\label{fig5}
\vfill
\end{figure}

\newpage
\clearpage
\begin{figure}[p]
\vfill
\setlength{\unitlength}{1.0cm}
\begin{picture} (18.0,18.0)
\put (0.0,0.0){\centerline{\epsfig{figure=\figdir 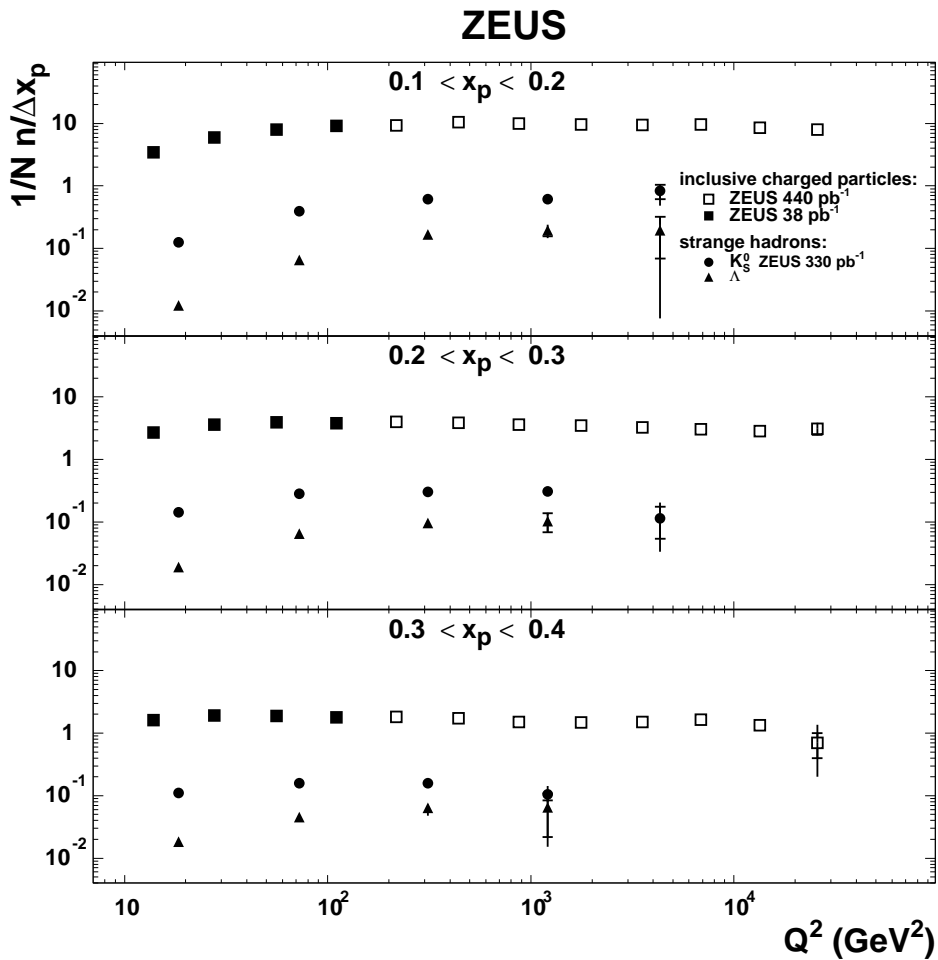,width=14cm}}}
\end{picture}
\caption
{\it 
The measured scaled momentum distributions $(1/N)(n(H)/\Delta x_p)$
for $H=\k0s$ (dots), $\Lambda$ (triangles) and light charged particles
(squares) as functions of $\q2$ in different regions of $x_p$. Other
details as in the caption to Fig.~\ref{fig2}.
}
\label{fig6}
\vfill
\end{figure}

\end{document}